\documentclass{aa}
\usepackage[numbers]{natbib}
\usepackage{psfig,graphicx,ulem}
\usepackage{txfonts}
\usepackage{soul,color}

\def\mathnew{\mathsurround=0pt}
\def\simov#1#2{\lower .5pt\vbox{\baselineskip0pt \lineskip-.5pt
\ialign{$\mathnew#1\hfil##\hfil$\crcr#2\crcr\sim\crcr}}}

\begin{document}

\title{The clusters Abell~222 and Abell~223: a multi-wavelength view
  \thanks{Based on observations obtained with MegaPrime/MegaCam, a
    joint project of CFHT and CEA/DAPNIA, at the Canada-France-Hawaii
    Telescope (CFHT) which is operated by the National Research
    Council (NRC) of Canada, the Institut National des Sciences de
    l'Univers of the Centre National de la Recherche Scientifique
    (CNRS) of France, and the University of Hawaii. This work has also
    made use of data products produced at TERAPIX and the Canadian
    Astronomy Data Centre as part of the Canada-France-Hawaii
    Telescope Legacy Survey, a collaborative project of NRC and CNRS.
    The X-ray analysis is based on XMM-Newton archive data.
    This research has made use of the NASA/IPAC Extragalactic Database
    (NED) which is operated by the Jet Propulsion Laboratory,
    California Institute of Technology, under contract with the
    National Aeronautics and Space Administration.}}

\author{
F.~Durret \inst{1,2} \and
T.~F.~Lagan\'a \inst{3} \and
C.~Adami \inst{4} \and
E.~Bertin  \inst{1,2} 
}

\institute{
UPMC Universit\'e Paris 06, UMR~7095, Institut d'Astrophysique de Paris, 
F-75014, Paris, France
\and
CNRS, UMR~7095, Institut d'Astrophysique de Paris, F-75014, Paris, France
\and
IAG, USP, R. do Mat\~ao 1226, 05508-090. S\~ao Paulo/SP, Brazil
\and
LAM, P\^ole de l'Etoile Site de Ch\^ateau-Gombert,
38 rue Fr\'ed\'eric Joliot-Curie,
13388 Marseille Cedex 13, France
}

\date{Accepted . Received ; Draft printed: \today}


\titlerunning{Abell 222/223}

\abstract
{The Abell 222 and 223 clusters are located at an average redshift
  z$\sim $0.21 and are separated by 0.26~deg. Signatures of mergers have been
  previously found in these clusters, both in X-rays and at optical
  wavelengths, thus motivating our study.  In X-rays, they are
  relatively bright, and Abell~223 shows a double structure. A
  filament has also been detected between the clusters both at optical
  and X-ray wavelengths.  }
{We analyse the optical properties of these two clusters based on deep
  imaging in two bands, derive their galaxy luminosity functions
  (GLFs) and correlate these properties with X-ray characteristics
  derived from XMM-Newton data.}
{The optical part of our study is based on archive images obtained
  with the CFHT Megaprime/Megacam camera, covering a total region of
  about 1~deg$^2$, or 12.3$\times$12.3~Mpc$^2$ at a redshift of
  0.21. The X-ray analysis is based on archive XMM-Newton images.}
{The GLFs of Abell~222 in the $g'$ and $r'$ bands are well fit by a
  Schechter function; the GLF is steeper in $r'$ than in $g'$. For
  Abell~223, the GLFs in both bands require a second component at
  bright magnitudes, added to a Schechter function; they are similar in both
  bands. The Serna \& Gerbal method allows to separate well the two
  clusters. No obvious filamentary structures are detected at very
  large scales around the clusters, but a third cluster at the same
  redshift, Abell~209, is located at a projected distance of
  19.2~Mpc. X-ray temperature and metallicity maps reveal that the
  temperature and metallicity of the X-ray gas are quite homogeneous
  in Abell~222, while they are very perturbed in Abell~223.}
{The Abell~222/Abell~223 system is complex. The two clusters that form
  this structure present very different dynamical states. Abell~222 is
  a smaller, less massive and almost isothermal cluster. On the other
  hand, Abell~223 is more massive and has most probably been crossed
  by a subcluster on its way to the northeast.  As a consequence, the
  temperature distribution is very inhomogeneous. Signs of recent
  interactions are also detected in the optical data where this
  cluster shows a ``perturbed'' GLF.  In summary, the multiwavelength
  analyses of Abell~222 and Abell~223 are used to investigate the
  connection between the ICM and the cluster galaxy properties in an
  interacting system.}

\keywords{Galaxies: clusters: individual (Abell~222, Abell~223),
  Galaxies: luminosity function}

\maketitle

\section{Introduction}

Galaxy evolution is known to be influenced by environmental effects,
which are particularly strong in galaxy clusters, all the more when
they are undergoing merging events. The analysis of galaxy luminosity
functions (GLF), and in particular of their faint-end slopes is a good
way to trace the history of the faint galaxy population and the
influence of merging events on this population. As summarized for
example in Table~A.1 of Bou\'e et al. (2008), this slope can strongly
vary from one cluster to another, and from one region to another in a
given cluster.

Superseding the dynamical analyses of Proust et al. (2000) and Girardi
\& Mezzetti (2001), Dietrich et al. (2002) have obtained a redshift
catalogue for 183 galaxies in the region of the cluster pair
Abell~222/223, out of which 153 belong to one of the two
clusters. Their analysis was also based on photometry in the V and R
bands.  They have estimated redshifts of 0.2126$\pm$0.0008
and 0.2079$\pm$0.0008, and velocity dispersions of 1014 and
1032~km~s$^{-1}$, for Abell~222 and Abell~223 respectively. As
indicated by Dietrich et al. (2002) these velocity dispersions are
somewhat higher than those expected from the X-ray
luminosities. Abell~222 appears relatively relaxed, with a galaxy
velocity histogram consistent with a gaussian. On the other hand, the
adaptive kernel galaxy density map of Abell~223 shows two peaks,
though surprisingly the Dressler-Schectman test (Dressler \&
Schectman 1988) does not find these two peaks and the DIP statistics
cannot reject the null hypothesis of a unimodal distribution, implying
that a Gaussian parent population cannot be excluded.  Dietrich et
al. (2002) derived respective mass to light ratios of 202$\pm$43 and
149$\pm$33 h$_{70}$ $M_\odot/L_\odot$ and luminosity functions in the
R band down to absolute magnitudes of $-19.5$, and found that a bridge
of galaxies seems to exist between the two clusters. Note that such
bridges of galaxies joining two clusters are still quite rare, another
example at a comparable redshift being the filament detected between
Abell~1763 and Abell~1770 by Fadda et al. (2008).

A weak lensing analysis of the Abell~222/223 system was performed by
Dietrich et al. (2005), who found evidence for a possible dark matter
filament connecting both clusters. They interpret the difference in
redshift between Abell~222 and Abell~223 as implying a separation
along the line of sight of 15$\pm 3$ h$_{70}^{-1}$~Mpc. Their weak
lensing surface mass density contours again show evidence for a double
structure in Abell~223.

The existence of a 1.2~Mpc wide filament between both clusters was
also found in X-rays by Werner et al. (2008).

We present here a muti-wavelength optical and X-ray analysis of the
clusters Abell~222 and Abell~223. The optical part is based on imaging
in the $g'$ and $r'$ bands obtained with the CFHT Megaprime/Megacam
camera and retrieved from the Megapipe image stacking pipeline at CADC
(Gwyn 2009). As seen in Table~\ref{tab:obs}, the $r'$ band data are
about 0.55~magnitude deeper than the $g'$ band data (by taking an
average color $g'-r'=0.85$), and will be given priority in the
morphological analysis; the $g'$ data will be used to select bright
cluster members from color-magnitude diagrams, and also because the
$g'$ band is more sensitive to star formation than the $r'$ band.
These images are deeper than those of Dietrich by at least 0.5~mag
(our raw galaxy counts start decreasing for $r'>24.5$ while Dietrich
et al. say that their counts in the R band stop following a power law
for R$>24$) , and therefore allow to probe the GLF to somewhat fainter
magnitudes. Our X-ray analysis is based on archive XMM-Newton data.

We will consider hereafter that the two clusters have a common
redshift of 0.21, for which Wight's cosmology calculator (Wright 2006)
gives a distance of 1035~Mpc, a spatial scale of 3.427~kpc/arcsec and
a distance modulus of 40.07 (assuming a flat $\Lambda$CDM cosmology
with H$_0=70$~km~s$^{-1}$~Mpc$^{-1}$, $\Omega _M=0.3$ and $\Omega
_\Lambda =0.7$).

The paper is organised as follows. We describe our optical analysis in
Section 2, and results concerning the galaxy luminosity function in
Section 3. A search for substructures in both clusters based on the
Serna-Gerbal method is described in Section~4.  The X-ray data
analysis and results, including temperature and metallicity maps, are
presented in Section~5.  An overall picture of this cluster pair is
drawn in Section~6.

\section{Optical data and analysis}

\subsection{The optical data}

We have retrieved from the CADC Megapipe archive (Gwyn 2009) the
reduced and stacked images in the $g'$ and $r'$ bands (namely
G002.024.493-12.783.G and G002.024.493-12.783.R) and give a few
details on the observations in Table~\ref{tab:obs}.  Observations were
made at the CFHT with the Megaprime/Megacam camera, which has a pixel
size of $0.186\times 0.186$~arcsec$^2$.

\begin{table}
\caption{Summary of the observations.}
\begin{center}
\begin{tabular}{lrr}
\hline
Filter          & $g'$    & $r'$ \\     
\hline
Number of coadded images & 5   & 8 \\
Exposure time (s)   & 2000  & 4000  \\
Seeing (arcsec)     & 0.81  & 0.73 \\
Limiting magnitude (5$\sigma$) & 26.7 & 26.4 \\
\hline
\end{tabular}
\end{center}
\label{tab:obs}
\end{table}

We did not use the catalogues available for these images, because they
were made without masking the surroundings of bright stars, so we
preferred to build masks first, then to extract sources with
SExtractor (Bertin \& Arnouts 1996).  Due to the dithering pattern,
the total area covered by the $r'$ image was slightly larger than the
Megacam 1~deg$^2$ field: 1.196~deg$^2$.

Objects were detected and measured in the full $r'$ image, then measured
in the $g'$ image in double image mode. Magnitudes are in the AB
system. The objects located in the masked regions were then taken out
of the catalogue, leading to a final catalogue of 223,414 objects. All
these objects have measured $r'$ magnitudes, and 214,669 also have
measured $g'$ magnitudes.

Since the $r'$ catalogue is deeper, and the seeing is also better in
this band (see Table~\ref{tab:obs}), we will perform our star-galaxy
separation in the $r'$ band.

\subsection{Star-galaxy separation}

\begin{figure} 
\centering \mbox{\psfig{figure=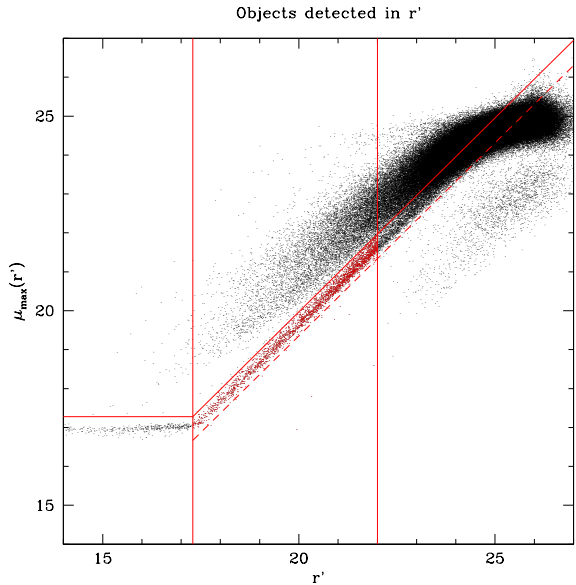,width=8cm}}
\caption[]{Central surface brightness in the $r'$ band as a function
  of $r'$ magnitude. The horizontal and oblique full red lines
  separate the galaxies (above these lines) from the stars (below
  these lines). The oblique dotted red line shows the limit for stars 
  computed for $17.25<r'<22$ (see text).  The two vertical lines
  correspond to $r'=17.25$ where the stars stop being saturated, and
  $r'=22$, where the fit to calculate the star sequence was limited
  (see text).}
\label{fig:r_mumax}
\end{figure}

In order to separate stars from galaxies, we plotted the maximum
surface brightness $\mu_{max}$ in the $r'$ band as a function of
$r'$. The result is shown in Fig.~\ref{fig:r_mumax}.

The best fit to the star sequence visible on Fig.~\ref{fig:r_mumax}
calculated for $17.25<r'<22$ is $\mu_{max}=0.996*r'-0.247$, with 
standard deviations on the slope and constant of 0.002 and 0.047
respectively. The point-source (hereafter called ``star'') sequence is
clearly visible for $r'<22$, with the star saturation showing well for
$r'<17.25$.  We will define galaxies as the objects: with $\mu
_{max}(r')>17.25$ for $r'<17.25$, and as the objects above the line of
equation $\mu_{max}=0.996*r'-0.247+0.3$, or
$\mu_{max}=0.996*r'+0.053$.  Stars will be defined as all the other
objects (see Fig.~\ref{fig:r_mumax}).  The small cloud of points
observed in Fig.~\ref{fig:r_mumax} under the star sequence is in fact
defects, but represents less than 2\% of the number of stars.  We thus
obtain a star and a galaxy catalogue.

As a check to see up to what magnitude we could trust our star-galaxy
separation, we retrieved the star catalogue from the Besan\c{c}on
model for our Galaxy (Robin et al. 2003) in a 1~deg$^2$ region
centered on the position of the image analysed here. Such a catalogue
is in AB magnitudes (as ours) and is corrected for extinction. In
order for it to be directly comparable to our star catalogue, we
corrected our star catalogue (and our galaxy catalogue as well, for
later purposes) for extinction: 0.071~mag in $g'$ and 0.052~mag in
$r'$ (as derived from the Schlegel et al. 1998 maps).

The $r'$ magnitude histogram of the objects classified as stars in our
$r'$ image roughly agrees with the Besan\c con star catalogue for
$r'\leq$20, though in some bins our counts tend to be somewhat higher
than those of the Besan\c con model even for bright magnitudes.  For
$r'>$20, we detect more stars than predicted by the Besan\c con model
(the difference is about 35\% at $r'\sim 21$).
This could imply that our star-galaxy separation is not valid for
$r'>$20; however, as suggested by Fig.~\ref{fig:r_mumax}, this is a
rather bright limit, as confirmed by simulations adapted to match
comparable data, where we found that the star-galaxy separation
determined with this method was reliable at least up to $r'=21$
(Bou\'e et al. 2008). In reality, the disagreement between our star
counts and those predicted by the Besan\c con model is most probably
due to the fact that our field is located in the direction of one of
the densest regions of the Sagittarius stream (Ibata et al. 2001,
Yanny et al. 2009). It is therefore normal to observe more stars than
predicted by the Besan\c con model, which does not take this stream
into account. Note that there may also be a small contribution due to
quasars and AGN, since their number at $r'\sim 20.5$ is expected to be
about 20 per square degree (see Richards et al. 2006, figure~13).

For $r'>20$, we will conservatively consider that the star-galaxy
separation may be wrong, and we will compute galaxy counts by counting
the total number of objects (galaxies plus stars) per bin of 0.5~mag,
and considering that the number of galaxies is equal to the total
number of objects minus the number of stars predicted in each bin by
the Besan\c con model. As mentioned below, we have checked that the
galaxy number counts in the $20<r'<22$ range estimated by both methods
(i.e. our star-galaxy separation and the subtraction of the Besan\c
con counts to the total number of objects) are consistent within error
bars.

\begin{figure} 
\centering \mbox{\psfig{figure=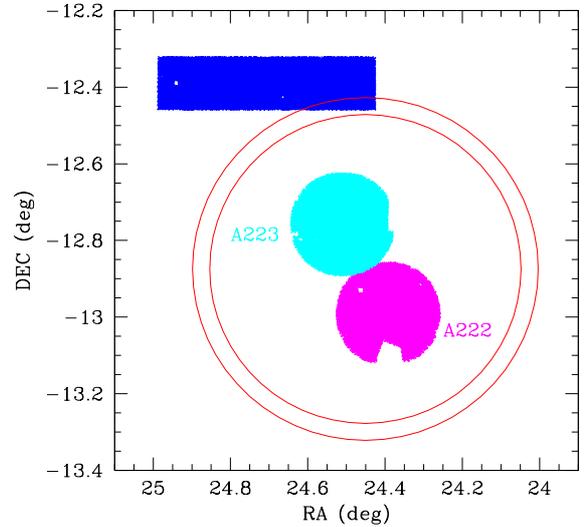,width=8cm}}
\caption[]{Positions of the objects (stars and galaxies) in the
  Abell~222 (magenta) and Abell~223 (cyan) catalogues. The objects
  used to estimate background counts (see text) are shown in blue for
  the north rectangle and in red for the annulus surrounding the
  clusters. Note that the figure covers $1.2\times 1.2$~deg$^2$,
  and is slightly larger than the images, which cover $1.04
    \times 1.15$~deg$^2$.}
\label{fig:xy}
\end{figure}

Since the clusters are quite distant, they do not cover the whole
field, so we extracted from the star and galaxy catalogues two
catalogues as large as possible corresponding to the two clusters.  We
defined the J2000.0 positions of Abell~222 and Abell~223 as coinciding
with the brightest galaxy of each cluster, i.e.  24.3921,$-12.9912$
and 24.5099,$-12.7575$ respectively (in degrees). Note that these
are not exactly the cluster positions given by the NED database. The
maximum possible radius to obtain independent catalogues for the two
clusters was 0.1290~deg, or 1.6~Mpc at a redshift of 0.21.

For the Abell~222 and Abell~223 clusters separately, we thus obtained
three complementary catalogues (with $g'$ and $r'$ magnitudes):
objects classified as galaxies (classification valid at least for
$r'\leq 20$), objects classified as stars, and a complete catalogue of
galaxies+stars which will be used for $r'>20$. The positions of the
galaxies in the regions of the two clusters are shown in
Fig.~\ref{fig:xy}.

\subsection{Catalogue completeness}

\begin{figure} \centering
\mbox{\psfig{figure=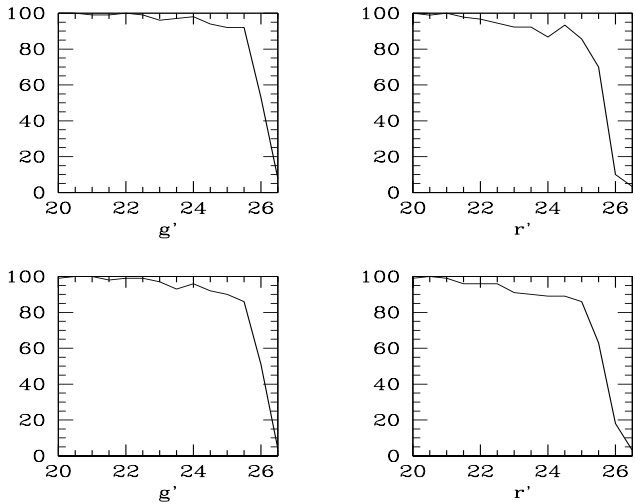,width=9cm,height=7cm,angle=0}}
\caption[]{Point source completeness as a function of magnitude in
  percentages for Abell~222 (top) and Abell~223 (bottom) in $g'$
  (left) and $r'$ (right) for point--like objects (see
  text). }  \label{fig:compl}
\end{figure}

The completeness of the catalogue is estimated by simulations.  For
this, we add ``artificial stars'' (i.e. 2-dimensional Gaussian
profiles with the same Full-Width at Half Maximum as the average image
Point Spread Function) of different magnitudes to the CCD images and
then attempt to recover them by running SExtractor again with the same
parameters used for object detection and classification on the
original images. In this way, the completeness is measured on the
original images.

In practice, we extract from the full field of view two subimages,
each $1500\times 1500$~pixels$^2$, corresponding to the positions of
the Abell~222 and Abell~223 clusters on the image.

In each subfield, and for each 0.5 magnitude bin between $r'=20$ and 27,
we generate and add to the image one star that we then try to detect
with SExtractor, assuming the same parameters as previously. This
process is repeated 100 times for each of the two fields and bands.

Such simulations give a completeness percentage for stars. This is
obviously an upper limit for the completeness level for galaxies,
since stars are easier to detect than galaxies. However, we have shown
in a previous paper that this method gives a good estimate of the
completeness for normal galaxies if we apply a shift of $\sim 0.5$~mag
(see Adami et al. 2006). Results are shown in Fig.~\ref{fig:compl}.

From these simulations, and taking into account the fact that results
are worse by $\sim 0.5$~mag for mean galaxy populations than for
stars, we can consider that our galaxy catalogue is complete to
better than 80\% for $g'\leq 25$ and $r'\leq 24.5$.

\subsection{Galaxy counts}

The surfaces covered by the Abell~222 and 223 catalogues (after
excluding masked regions) are 0.0429~deg$^2$ and 0.0475~deg$^2$
respectively. Galaxy counts were computed in bins of 0.5~mag
normalized to a surface of 1~deg$^2$.

For $r' \leq20$, galaxy counts were derived directly by computing
histograms of the numbers of galaxies in the Abell~222 and Abell~223
catalogues.  For $r'>20$, we built for each cluster histograms of the
total numbers of objects (galaxies+stars) and obtained galaxy counts
by subtracting the numbers of stars predicted by the Besan\c con
model.  The resulting galaxy counts will be used in the next section
to derive the GLFs for both clusters in both bands.

As a test, we considered the galaxy counts in the four 0.5 magnitude
bins for $20<r'<22$ computed for both clusters with the two
methods (i.e.  first method: considering that the star-galaxy
  separation is valid, and second method: considering the total number
  of objects (galaxies+stars) and subtracting the number of stars
  predicted by the Besan\c con model to obtain the number of galaxies). 
In all cases, the differences are smaller than 15\% (and in most cases
they are only a few percent). Therefore, the limit (between $r'=20$
and 22) at which we consider that our star-galaxy separation is not
valid any more will not strongly influence our results.
 
Note that no k-correction was applied to the galaxy magnitudes.

\section{Results: colour-magnitude diagrams and galaxy luminosity
  functions}

In order to compute the galaxy luminosity functions of the two
clusters, it is necessary to subtract to the total galaxy counts the
number counts corresponding to the contamination by the foreground and
background galaxies.

For galaxies brighter than $r'=20$ we will select galaxies with a high
probability to belong to the clusters by drawing colour-magnitude
diagrams and selecting galaxies located close to this relation.  A few
spirals may be missed in this way, but their number in any case is
expected to be small, as explained at the end of Sect.~3.1 (also see
e.g. Adami et al. 1998).  For galaxies fainter than $r'=20$ we will
subtract galaxy counts statistically.

\subsection{Colour--magnitude diagrams}

\begin{figure} 
\centering \mbox{\psfig{figure=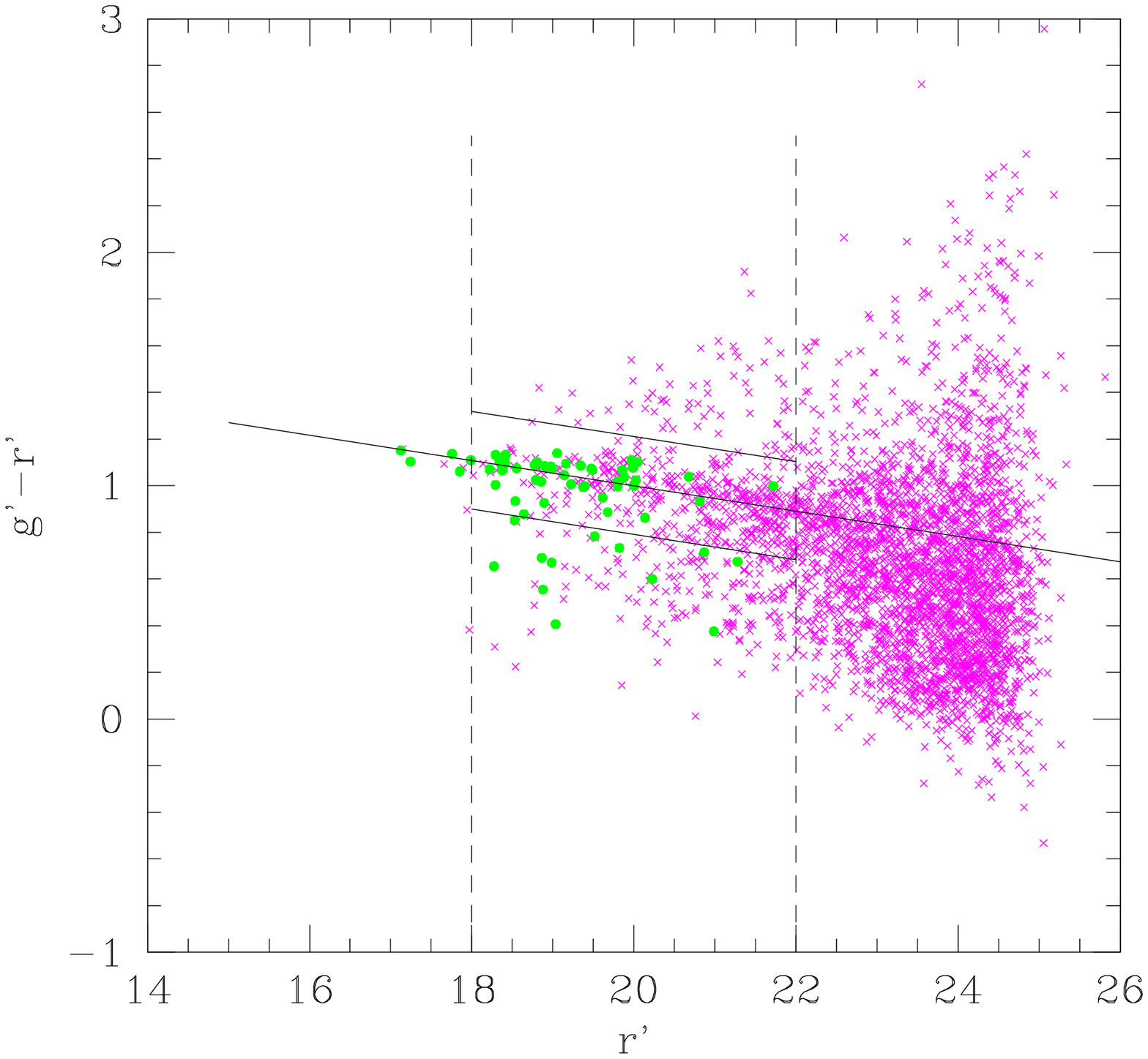,width=7cm}}
\centering \mbox{\psfig{figure=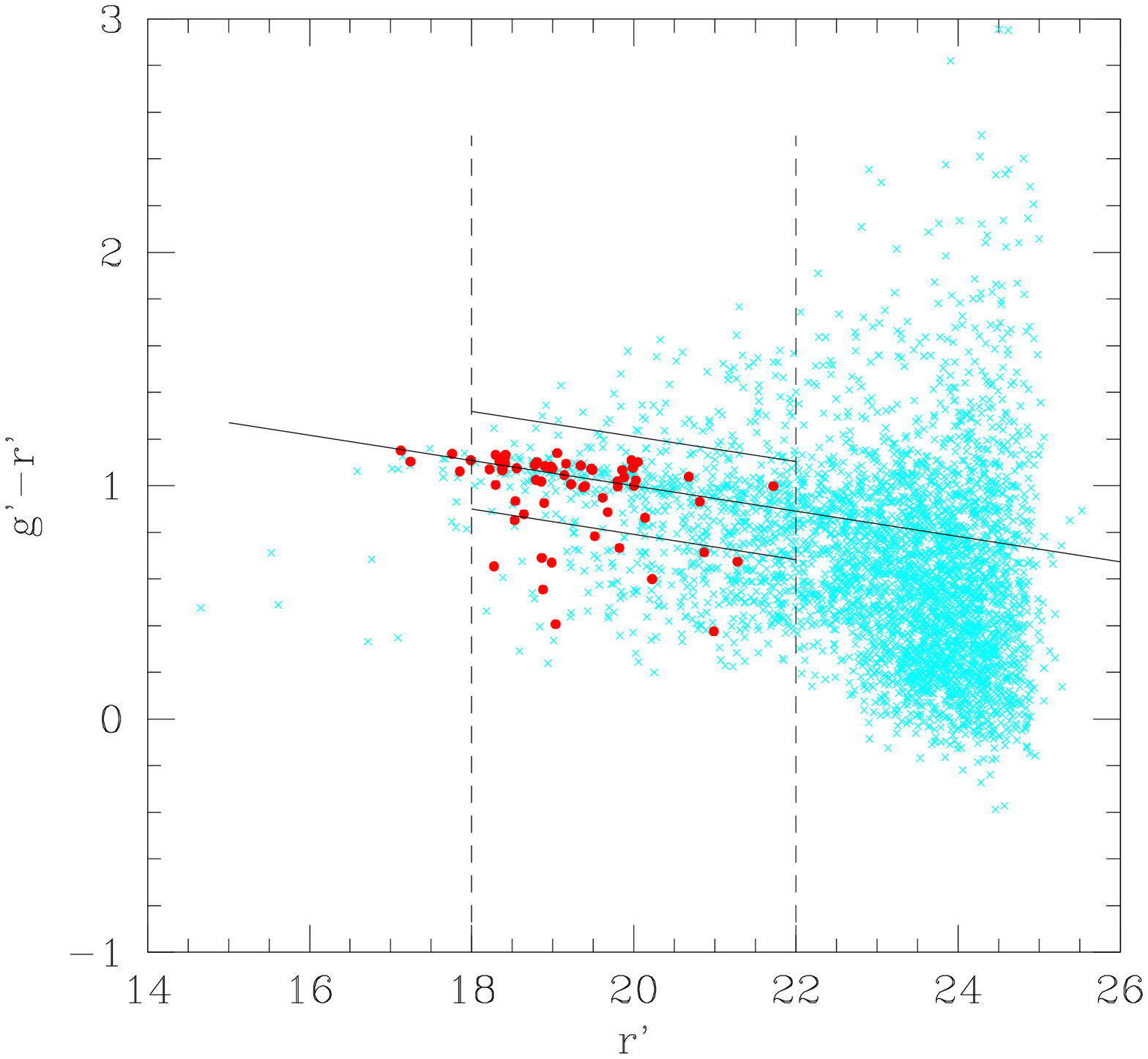,width=7cm}}
\caption[]{$(g'-r')$ vs. $r'$ colour--magnitude diagrams for Abell~222
  (top) and Abell~223 (bottom) for objects classified as galaxies from
  the $\mu_{max}$-magnitude relation. The vertical dashed lines
  indicate the magnitude interval where the colour-magnitude relation
  was computed. The long oblique line shows the colour-magnitude
  relation, while the two short oblique lines indicate an interval of
  $\pm 3\sigma$ around the colour-magnitude relation. The filled
  circles show the galaxies belonging to the clusters according to
  their spectroscopic redshifts. }
\label{fig:coulmag}
\end{figure}

The mean values for
$g'-r'$ are 0.85 and 0.87 for Abell~222 and Abell~223.
The $g'-r'$ vs. $r'$ colour--magnitude diagrams are shown in
Fig.~\ref{fig:coulmag} for the two clusters. A sequence is well
defined for galaxies in the magnitude range $18<r'<22$ in both
clusters. We computed the best fit to the $g'-r'$ vs. $r'$ relations
in this magnitude range by applying a linear regression (separately
for the two clusters, since their colour-magnitude relations may not
be absolutely identical). We then eliminated the galaxies located 
more than 3$\sigma$ away from this relation and computed the $g'-r'$
vs. $r'$ relations again.

The equations of the colour--magnitude relations are found to be:
$g'-r'=-0.0545r'+2.091$ with $3\sigma=0.21$ for Abell~222, and
$g'-r'=-0.0474r'+1.945$ with $3\sigma=0.21$ for Abell~223.

For $r'<20$, we will consider hereafter that all the galaxies located
within 3$\sigma$ of these relations (i.e. between the two black lines of
Fig.~\ref{fig:coulmag}) belong to the clusters. With this condition,
there are 141 and 144 galaxies belonging to Abell~222 and Abell~223
respectively with $r'\leq 20$.

We also plot in Fig.~\ref{fig:coulmag} the galaxies belonging to the
clusters according to their spectroscopic redshifts, i.e. galaxies in
the [0.195, 0.215] redshift interval (see Sect.~4). We can see that
their positions agree well with the colour-magnitude selection.

For both clusters, we have estimated the number of blue cluster
galaxies lost by selecting galaxies in the red sequence interval in
the following way. First, we computed histograms of numbers of
galaxies in the red sequence and below the red sequence in bins of 1
absolute magnitude in the $r'$ band. The bins of interest here are
${\rm M_{r'}=-21.5}$ and $-20.5$ (see Fig.~\ref{fig:coulmag}), roughly
corresponding to $r'=18.5$ and 19.5.  Then we estimated the number of
foreground galaxies expected.  Since the comoving volume at z=0.21 is
2.622 Gpc$^3$ (Wright 2006), and each of our clusters covers an area
of 0.0522~deg$^2$ on the sky, the volume in the direction of each
cluster is 3318~Mpc$^3$. By using the R band luminosity function by
Ilbert et al. (2005) in the [0.05-0.20] redshift bin (see their Fig. 6
and Table 1), we find that the percentages of ``lost'' galaxies are of
the order of 10\%--25\% for ${\rm M}_{r'}=-20.5$ and of 10\% for ${\rm
  M}_{r'}=-21.5$.

\subsection{Comparison field}
\label{sec:compfield}

In order to perform a statistical subtraction of the background
contribution for $r'>20$, we considered several possibilities. 

\begin{figure} 
\centering \mbox{\psfig{figure=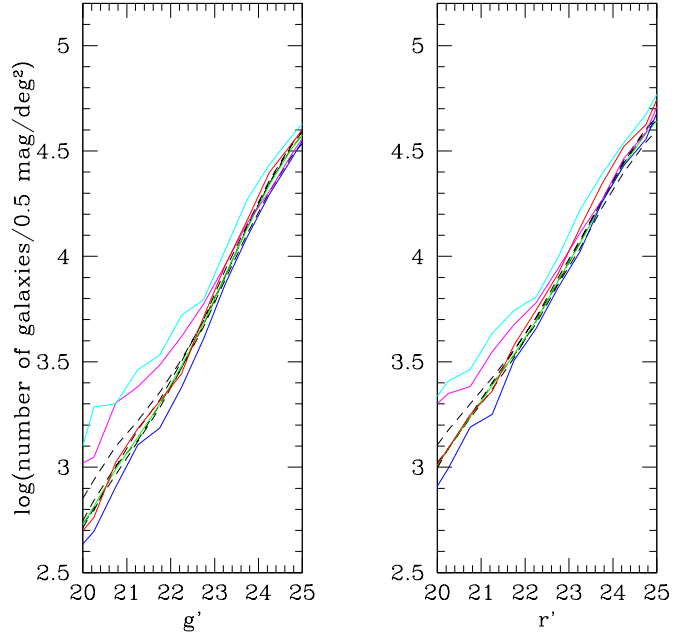,width=9cm}}
\caption[]{Galaxy counts in the $g'$ (left) and $r'$ (right) bands for
  magnitudes $r'>20$ where the background must be subtracted
  statistically, in logarithmic scale.  The counts in Abell~222 and
  223 are drawn in magenta and cyan respectively. The black
    dashed lines show the galaxy counts from the CFHTLS Deep survey,
    and the full green line shows the average of the Deep~1, Deep~2 and
    Deep~3 field counts.  The blue and red lines correspond to the
    ``local'' backgrounds extracted from the same images as our
    clusters (red: in the annulus between the two circles of
    Fig.~\ref{fig:xy}, blue: in the north rectangle, see text
    and Fig.~\ref{fig:xy}). Error bars are Poissonian and are not
  plotted for clarity.}
\label{fig:cts_bkgd}
\end{figure}

The most obvious solution would be to extract the background in an
annulus surrounding the clusters. The largest outer circle that can be
extracted, centered on the middle position between the two clusters
(24.4510, $-12.8744$ J2000.0 in degrees), has a radius of
0.4472~deg, or 5.55~Mpc. We can then take an inner circle of radius
0.4028~deg, or 5~Mpc. This annulus is drawn in Fig.~\ref{fig:xy}.
However, the number counts in this circle are higher than the CFHLTS
Deep survey counts (see below), probably because the annulus is too
close to the clusters and still includes cluster galaxies, so we will
not use these counts as a background.

The Canada-France-Hawaii Telescope Legacy Survey (CFHTLS) has been
taken with the same telescope, camera and filters as the data we are
analyzing. The Deep survey explores a solid angle of $4\times 1$~deg$^2$ of the
deep Universe, in four independent fields
(http://www.cfht.hawaii.edu/Science/CFHLS/).  Observations are carried
out in five filters ($u*,g',r',i'$ and $z'$) providing catalogs of
sources that are 80\% complete up to $i_{AB}$=26.0  (see
http://terapix.iap.fr/cplt/oldSite/Descart/CFHTLS-T0005-Release.pdf).
We only consider the Deep survey here, because we are interested in the
GLFs down to faint magnitudes.  For the four deep fields, we computed
the galaxy number counts in bins of 0.5 mag, in $g'$ and $r'$,
normalized to a surface of 1~deg$^2$. The galaxy counts in these four
fields are drawn in Fig.~\ref{fig:cts_bkgd}. As discussed before (see
e.g. Bou\'e et al. 2008 and references therein), the counts in these
four fields somewhat differ, due to cosmic variance, and in particular
the counts in the Deep~4 field are higher than in the other three. We
will therefore take as background galaxy counts the average of the
Deep~1, Deep~2 and Deep~3 field counts.

For comparison, we also extracted galaxy counts from a region assumed
to be representative of the background in our image. This region was
chosen to be a rectangle north of the clusters, in an area devoid of
bright stars. It is shown as a blue rectangle in Fig.~\ref{fig:xy} and
covers an effective area of 0.06825~deg$^2$.  The corresponding counts
(hereafter the ``local'' background counts) appear to be notably lower
than the CFHTLS-Deep counts (Fig.~\ref{fig:cts_bkgd}); this may be the
case if this region corresponds to a void located between the large
scale filaments that converge towards the clusters.

Although we consider that the CFHTLS-Deep counts probably represent
better the galaxy background counts, we will also fit the GLFs with
these ``local'' background counts, keeping in mind the fact that these
``local'' counts are likely to overestimate the GLFs thus derived for
Abell~222 and 223. This will illustrate the difficulty to determine
GLF parameters unambiguously.
 
We can note from Fig.~\ref{fig:cts_bkgd} that galaxy counts are
notably higher in Abell~223 than in Abell~222, implying that the
former cluster is richer than the latter.

\subsection{Galaxy luminosity functions}
\label{sec:glf}

The Galaxy Luminosity Functions (GLFs) of Abell~222 and Abell~223 were
calculated in bins of 0.5~mag and normalized to 1~deg$^2$, as
described above. We subtracted the background contribution using
  as background galaxy counts: 1)~the average of the three Deep~1,
  Deep~2 and Deep~3 fields; 2)~the ``local'' counts in the north blue
  rectangle of Fig.~\ref{fig:xy}.

\begin{figure} 
\centering \mbox{\psfig{figure=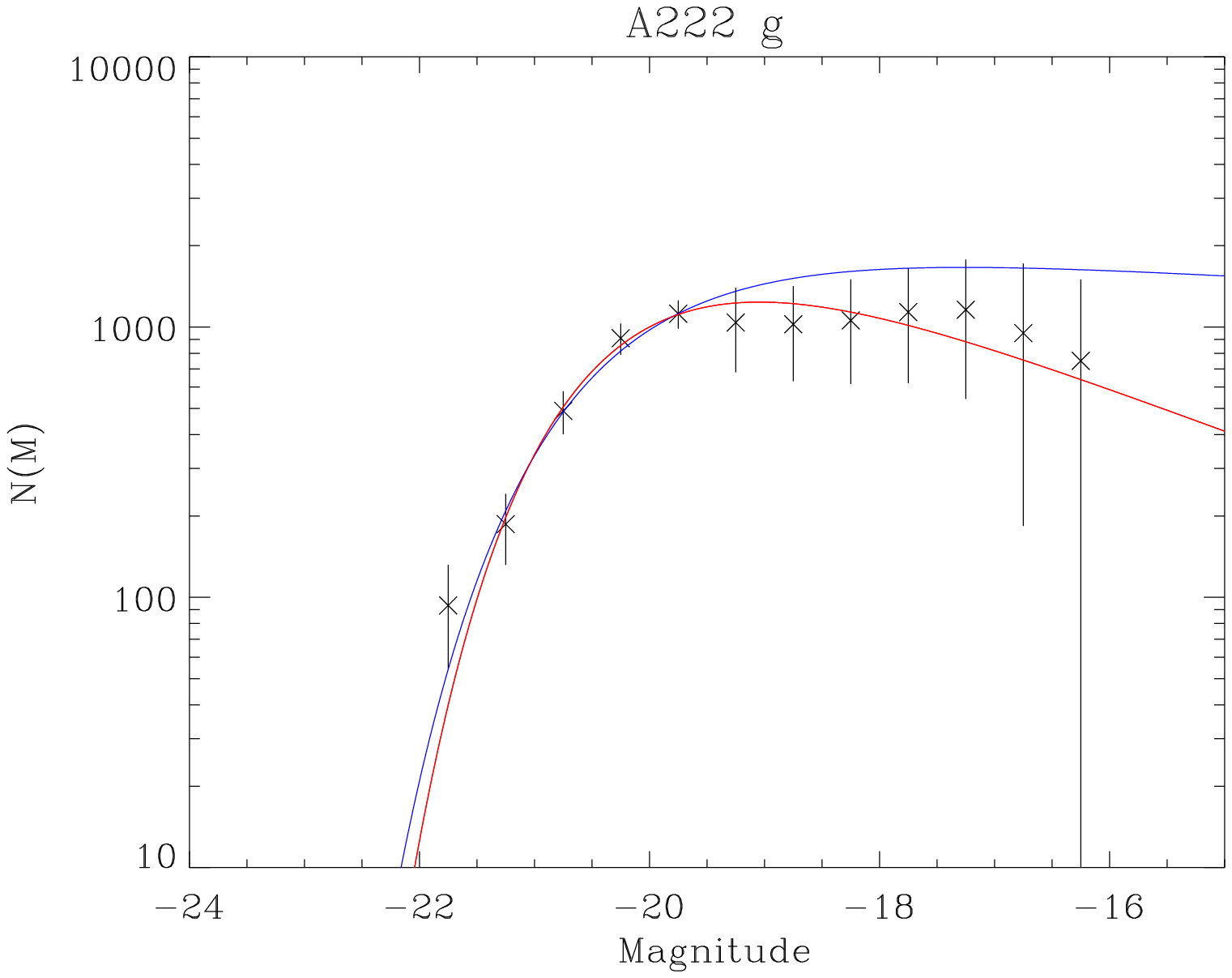,width=8cm}}
\centering \mbox{\psfig{figure=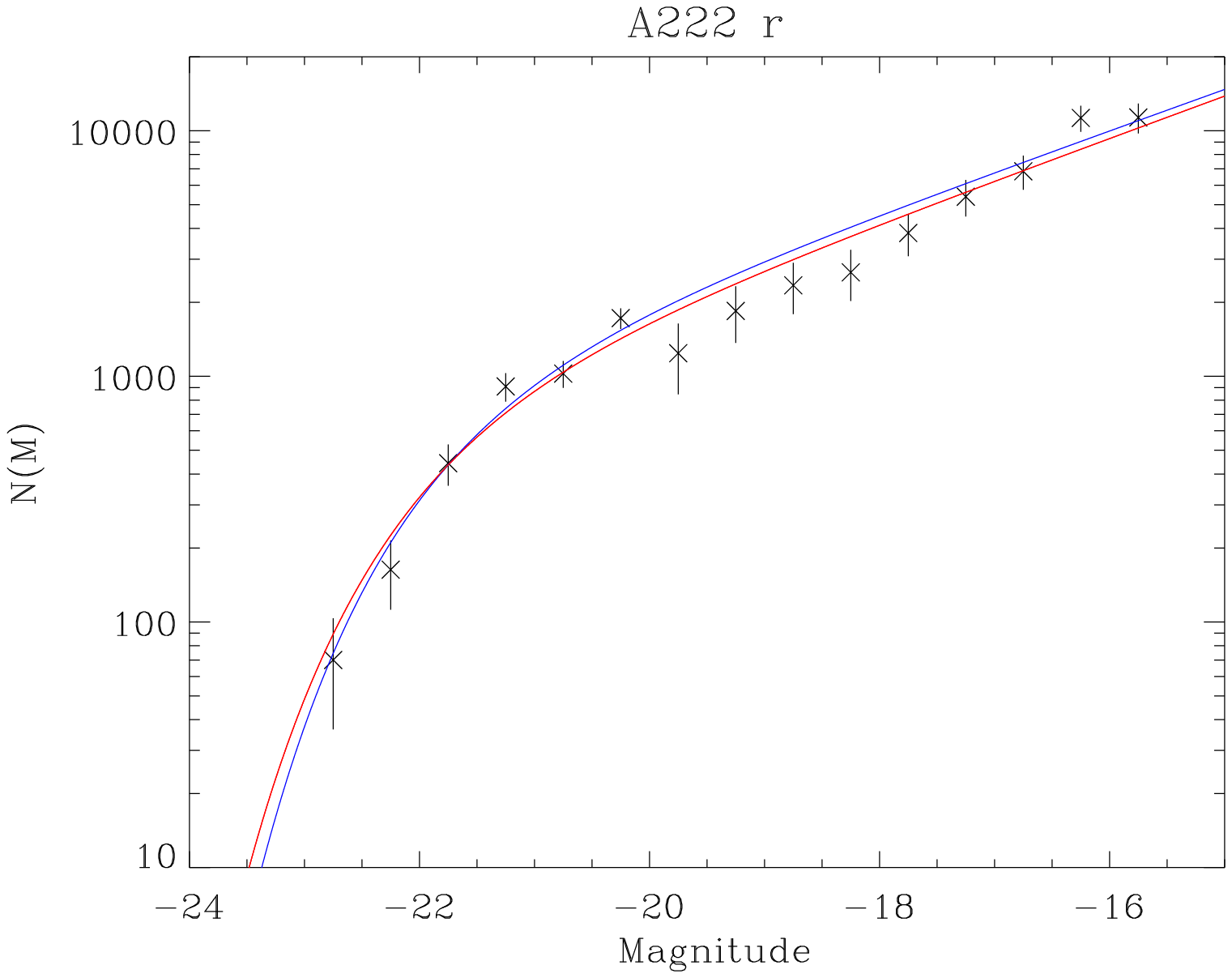,width=8cm}}
\caption[]{Galaxy luminosity functions for Abell~222 in the $g'$ (top)
  and $r'$ (bottom) bands, in logarithmic scale. The pooints
  correspond to the subtraction of the background counts taken from
  the CFHTLS. Error bars are 4 times the Poissonian errors on galaxy
  counts (see text). The best Schechter function fits are drawn
    in red for the subtraction of background counts taken from the
    CFHTLS and in blue for ``local'' background counts (see text).} 
\label{fig:GLF_A222}
\end{figure}

\begin{figure} 
\centering \mbox{\psfig{figure=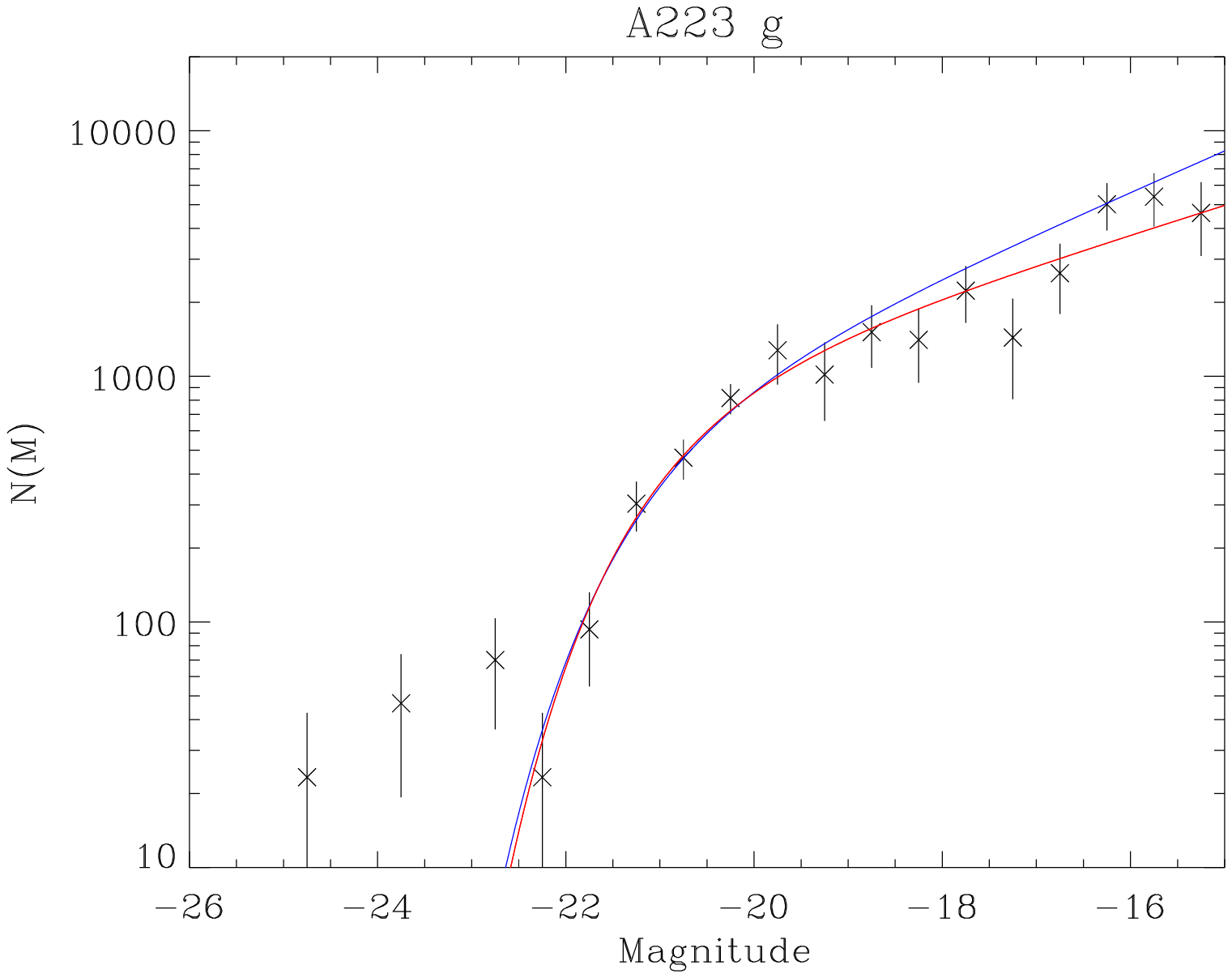,width=8cm}}
\centering \mbox{\psfig{figure=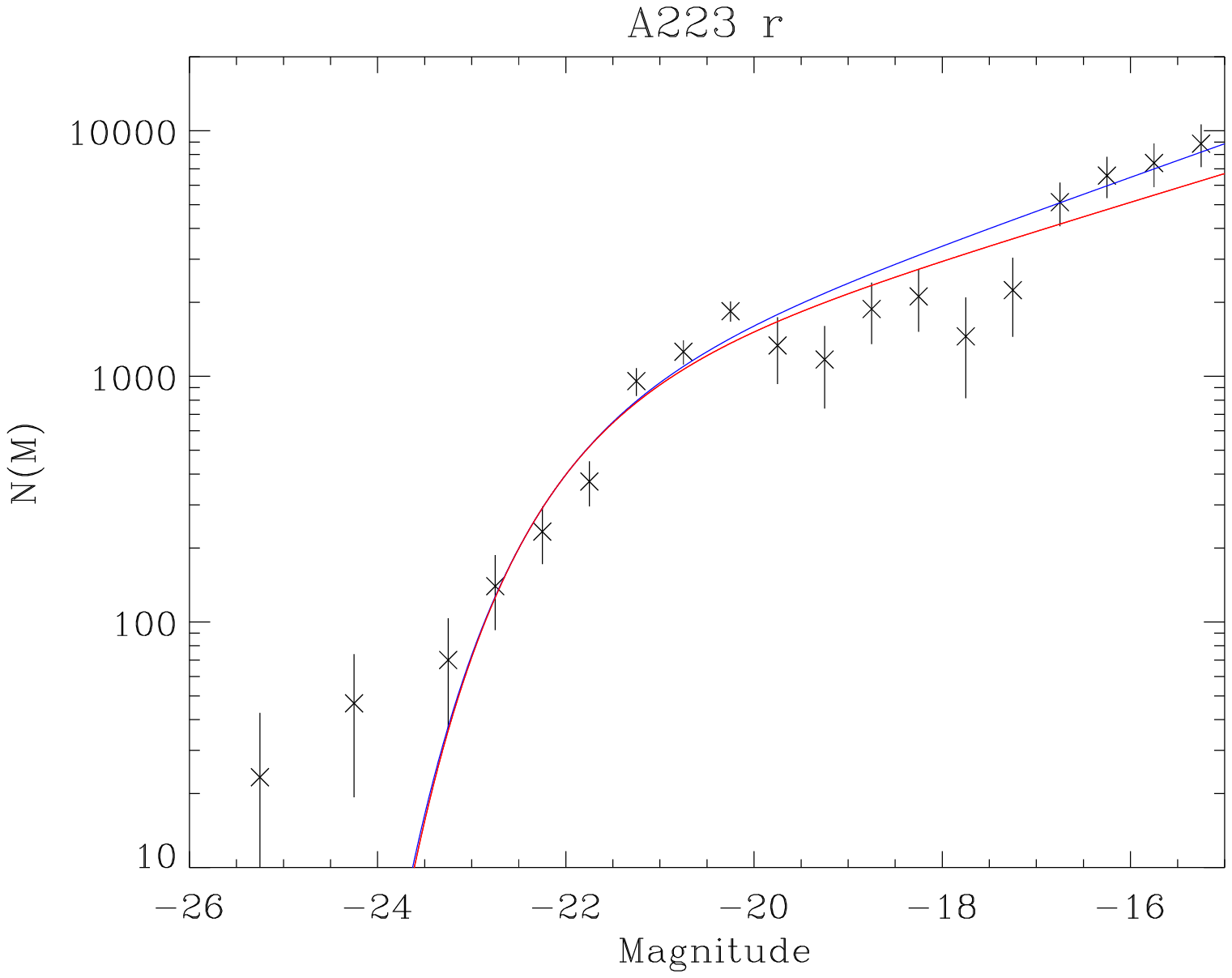,width=8cm}}
\caption[]{Same as Fig.~\ref{fig:GLF_A222} for Abell~223.  }
\label{fig:GLF_A223}
\end{figure}

The GLFs are displayed for Abell~222 and Abell~223 in
Figs.~\ref{fig:GLF_A222} and \ref{fig:GLF_A223} respectively.  The
error bars drawn in these figures were taken to be 4 times the
Poissonian errors on galaxy counts, as suggested by detailed
simulations previously performed by our team for similar data (see
Bou\'e et al. 2008, Fig.~5).

The GLFs (as a function of absolute magnitude) were fit by a Schechter function:

$$ S(M) = 0.4 \, \ln{10} \, \phi^{\ast} \, y^{\alpha+1} \, e^{-y} $$
with 
$y=10^{0.4 \, (M^{\ast}-M)} $.

The parameters of the Schechter function fits of the GLFs are
respectively given in Tables~\ref{tab:schechter_deep} and
\ref{tab:schechter_rect} for the two different background
subtractions: the average of the CFHTLS Deep fields and the ``local''
background. The absolute magnitude range considered is indicated for
each fit. The fits are drawn in Figs.~\ref{fig:GLF_A222} and
\ref{fig:GLF_A223} for Abell~222 and Abell~223 respectively (with
background galaxy counts taken from the CFHTLS Deep field counts).
 
\begin{table*}
\caption{Schechter parameters for galaxy luminosity functions
(background galaxy counts taken from the CFHTLS Deep field counts).}
\begin{center}
\begin{tabular}{lccccc}
\hline
Cluster   & Filter & Range & $\Phi ^*$ & M$^*$   & $\alpha$   \\     
\hline
Abell~222 & $g'$ & $[-22.5,-16.0]$ & $2891 \pm 134$ & $-20.04 \pm 0.06$ & $-0.60 \pm 0.05$ \\
          & $r'$ & $[-23.5,-16.0]$ & ~$937 \pm 64 $ & $-22.01 \pm 0.07$ & $-1.43 \pm 0.01$ \\
Abell~223 & $g'$ & $[-25.5,-15.5]$ & $1017 \pm 72 $ & $-21.05 \pm 0.06$ & $-1.30 \pm 0.02$ \\
          & $r'$ & $[-26.0,-15.0]$ & $1107 \pm 62 $ & $-22.05 \pm 0.06$ & $-1.29 \pm 0.01$ \\
\hline
\end{tabular}
\end{center}
\label{tab:schechter_deep}
\end{table*}

\begin{table*}
\caption{Schechter parameters for galaxy luminosity functions
(``local'' background galaxy counts).}
\begin{center}
\begin{tabular}{lccccc}
\hline
Cluster   & Filter & Range & $\Phi ^*$ & M$^*$   & $\alpha$   \\     
\hline
Abell~222 & $g'$ & $[-22.0,-16.0]$ & $2269 \pm 133$ & $-20.32 \pm 0.06$ & $-0.94 \pm 0.03$ \\
          & $r'$ & $[-23.0,-16.0]$ & $1132 \pm 63 $ & $-21.85 \pm 0.06$ & $-1.42 \pm 0.01$ \\
Abell~223 & $g'$ & $[-25.0,-15.0]$ & ~$819 \pm 55 $ & $-21.20 \pm 0.05$ & $-1.42 \pm 0.01$ \\
          & $r'$ & $[-26.0,-15.0]$ & $1042 \pm 58 $ & $-22.10 \pm 0.06$ & $-1.34 \pm 0.01$ \\
\hline
\end{tabular}
\end{center}
\label{tab:schechter_rect}
\end{table*}

If we compare the results given in Tables~\ref{tab:schechter_deep} and
\ref{tab:schechter_rect}, we see that the $\Phi ^*$ and M$^*$
parameters depend little on the background galaxy counts chosen.  On
the other hand, some variations are found in the faint end slope
$\alpha$, as expected since it is at faint magnitudes that the
background galaxy subtraction has a strong influence.

For Abell~222, the GLF is quite well fit by a Schechter function in
the $r'$ band, with a faint end slope $\alpha=-1.43$ (the same for
both background subtractions).  On the other hand, the GLF is
surprisingly flat in $g'$ (slope $\alpha=-0.60$ or $-0.94$, depending
on background subtraction).

A Schechter function is obviously not sufficient to fit the GLFs of
Abell~223, both in $g'$ and $r'$: Abell~223 has more very bright
galaxies than Abell~222, and a second component at bright magnitudes
is obviously required. At faint magnitudes, the GLF slopes in $g'$ and
$r'$ are comparable for Abell~223. They vary by small quantities (0.12
and 0.05 respectively) between one galaxy background subtraction and
the other.  However, these slopes are only indicative, since as seen
in Fig.~\ref{fig:GLF_A223} the GLFs show strong wiggles for absolute
magnitudes between $-20$ and $-17$ in $r'$ and between $-20$ and $-16$
in $g'$.  Since Abell~223 is itself a double cluster (Dietrich et
al. 2002), it is not surprising to find that it has a ``perturbed''
GLF. This is also the case for example for the Coma cluster, known to
have a two-component GLF (Biviano et al. 1995).

If we compare the GLFs of the two clusters, we can see that they are
more or less comparable in the $r'$ band (except for the wiggles in
Abell~223), while in the $g'$ band the GLF is notably steeper in
Abell~223 than in Abell~222. An explanation for this can be that both
clusters have comparable old galaxy populations, but that the number of
faint star forming galaxies is higher in Abell~223, where star
formation can be triggered by recent interactions.

If we compare these GLFs to those found by other authors, we can note
that the bright part of the GLF Schechter fits ($r'< 18$, or ${\rm
  M_{r'}<-18.9}$) for both clusters are very similar in shape to the
GLFs recently obtained e.g. by Andreon et al. (2008); these authors
analyzed the GLFs of a sample of clusters at various redshifts,
limited to relatively bright absolute magnitudes: ${\rm M_V}<-19$.

The $\alpha$ slopes of the faint ends of the GLFs derived here (except
for that of Abell~222 in the $g'$ band) are within the broad range of
values estimated by previous authors for different clusters, cluster
regions and photometric bands (see e.g. the compilation in Table~A1 of
Bou\'e et al. 2008).

Note that Abell~222 and Abell~223 are at redshift 0.21, and few GLFs
are available for clusters at such redshifts.  Andreon et al. (2005)
found somewhat shallower slopes for three clusters at redshifts $\sim
0.3$, but in the K band, so the comparison with our results is not
straightforward.

\subsection{Morphological segregation}
\label{subsec:morpho}

\begin{figure} 
\centering \mbox{\psfig{figure=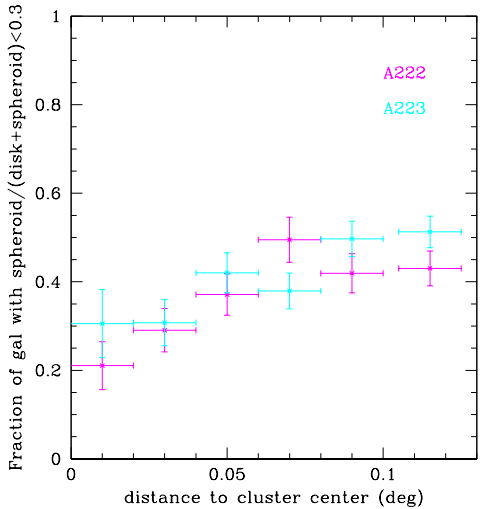,width=7cm}}
\caption[]{Fraction of disk-dominated galaxies as a function of
  radius for Abell~222 (in magenta) and Abell~223 (in cyan). }
\label{fig:ntyperay}
\end{figure}

\begin{figure} 
\centering \mbox{\psfig{figure=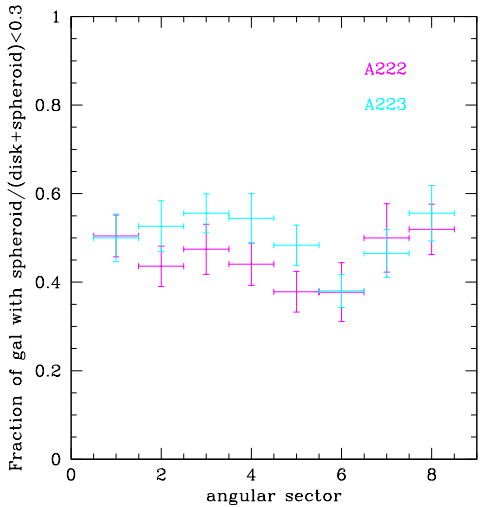,width=7cm}}
\caption[]{Fraction of disk-dominated galaxies in sectors of 45~deg
  within each cluster, with the same colours as in the previous
  figure. Angular sectors are numbered from 1 to 8 clockwise from
  east (see Fig.~\ref{secteurs}).}
\label{fig:ntypesec}
\end{figure}

We applied a new tool developed in SExtractor that calculates for each
galaxy the respective fluxes in the bulge (spheroid) and
disk. This new experimental {\sc SExtractor} feature fits to
each galaxy a two-dimensional model comprised of a de Vaucouleurs
spheroid (bulge) and an exponential disk. Briefly, the fitting process
is very similar to that of the GalFit package (Peng et al. 2002) and
is based on a modified Levenberg-Marquardt minimisation algorithm. The
model is convolved with a supersampled model of the local Point Spread
Function (PSF), and downsampled to the final image resolution.  The
PSF model used in the fits was derived with the {\sc PSFEx} software
(Bertin et al. 2010, in preparation) from a selection of 5436 ($g'$) /
6922 ($r'$) point source images. The PSF variations were fit using
a 6th degree polynomial of $x$ and $y$ image coordinates.

The model fitting was carried out independently in the $g'$ and $r'$
bands. The bulge to total ratios discussed here were extracted from
the red channel, limited to $r'=22$, since beyond this magnitude
results may become unreliable.

We applied this tool to look for evidence for morphological
segregation, by computing the fraction of galaxies with a spheroid to
total flux smaller than 0.3, as a function of radius (i.e. in
concentric circles of step 0.02~deg). The result is plotted in
Fig.~\ref{fig:ntyperay}.  In each bin, 1$\sigma$ errors were computed
as follows: if n is the number of galaxies having a spheroid to total
flux smaller than 0.3 and N the total number of galaxies, we assume
that n and (N-n) are independent variables; in this case, the error on
n/N is:
$$   (n/N)\ \sqrt{(1/n - 1/N)}. $$
We can see that there is a general increase of the fraction of
disk-dominated galaxies with radius as expected (see e.g. Biviano et
al. 1997).

In order to see if there was a morphological segregation that could be
linked to the presence of cosmological filaments, we computed that
same fraction in angles of 45~deg all around each cluster. The result
is shown in Fig.~\ref{fig:ntypesec}, where angular sectors are
numbered from 1 to 8 clockwise from east (see Fig.~\ref{secteurs}).
No strong trend is found in this plot within error bars, as expected
if there is no obvious cosmological filament linking the clusters with
the surrounding cosmic web. We can note however that for both clusters
there is a dip in the fraction of bulge-dominated galaxies in
sector~6, which corresponds to the line joining the two clusters; the
filament already detected by Werner et al. (2008) using
wavelet-decomposition in the [0.5-2]~keV energy band is located along
this same direction. In this zone the gas is not yet hot enough to
influence the galaxies. If we look at figure 9 from Lagan\'a et
al. (2009), 2~keV is the lower limit where ram-pressure plays an
important role in galaxy clusters. Thus we would expect to detect a
trend only if the filament was hotter than 2~keV.

Another possibility for this dip in the fraction of bulge-dominated
galaxies is a higher number of blue galaxies in which star formation
was triggered due to ongoing collisions.  Poggianti et al. (2004)
argued that there is a striking correlation between the positions of
the young and strong post-starburst galaxies and substructure in the
hot intra-cluster medium (ICM) identified from XMM-Newton data, with
these galaxies lying close to the edges of two infalling
substructures.  This result suggests that the interaction with the
dense ICM could be responsible for the triggering of the star
formation.

\subsection{Luminosity segregation}

\begin{figure} 
\centering \mbox{\psfig{figure=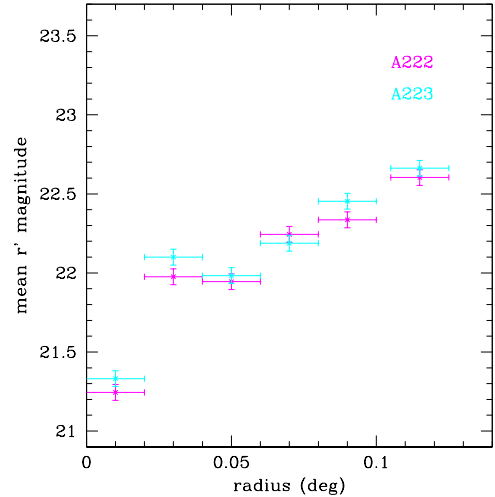,width=7cm}}
\caption[]{Average $r'$ magnitude as a function of
  radius for Abell~222 (in magenta) and Abell~223 (in cyan). }
\label{fig:magray}
\end{figure}

We searched for luminosity segregation in both clusters by selecting
galaxies brighter than $r'=24$ and within the red sequence defined in
Fig.~\ref{fig:coulmag}, to increase the probability that they belong
to the clusters. The mean $r'$ magnitude was calculated in each of the
radial bins previously defined and results are shown in
Fig.\ref{fig:magray}. We observe an increase of the mean magnitude
with radius: it is about 21.3 in the central bin, and then increases
with radius from values around 22 to 22.6.

\section{Cluster-scale to large-scale structure}

We searched the NED database for galaxies with redshifts available in
a very large region of 3~deg radius around Abell~222/223 and found 980
galaxies.  This radius corresponds to 37~Mpc at the distance of the
studied clusters, i.e. to the typical diameter of a ``bubble'' or of a
void in the universe (see e.g. Hoyle $\&$ Vogeley 2004).  This
catalogue of galaxies with measured redshifts will be used first to
search for substructures in the zones covered by our clusters, and
second to search for structures (such as filaments) at a larger scale.

\subsection{Search for gravitationally bound structures}

\begin{figure} 
\centering \mbox{\psfig{figure=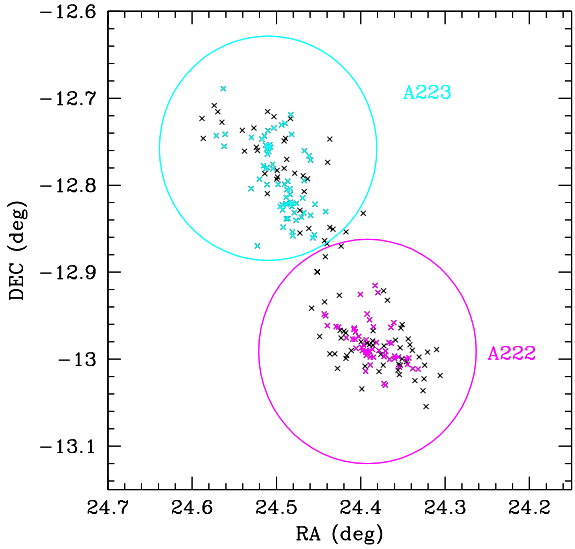,width=8cm}}
\caption[]{Positions of the galaxies with measured redshifts and
  magnitudes in the regions covered by Abell~222 and Abell~223 (black
  crosses). The galaxies belonging to the two main gravitationally
  bound systems identified with the SG method are shown in colours
  (see text). }
\label{fig:xy_z}
\end{figure}

\begin{figure} 
\centering \mbox{\psfig{figure=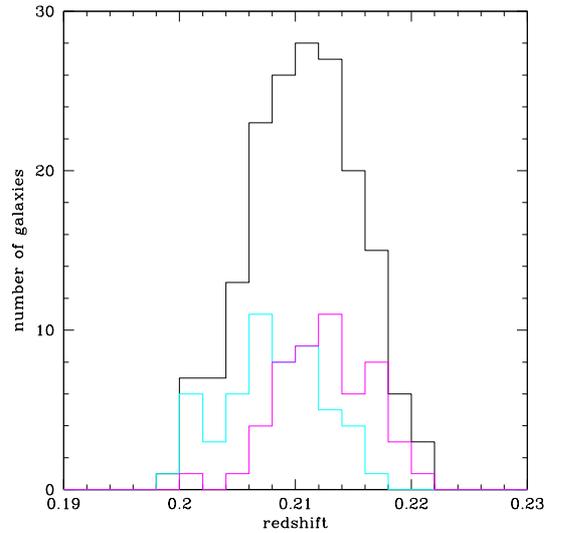,width=8cm}}
\caption[]{Redshift histograms for all the galaxies with measured
  redshifts in the regions covered by Abell~222 and Abell~223 (in
  black), and for the galaxies in the gravitationally bound structures
  of Abell~222 and Abell~223, in magenta and cyan respectively (the
  positions of these galaxies are shown in Fig.~\ref{fig:xy_z}) with
  the same colour coding. }
\label{fig:histoz}
\end{figure}

Within the regions of our images covered by Abell~222 and Abell~223,
246 galaxies have measured redshifts, and 210 galaxies have both
measured redshifts {\sl and} magnitudes. Out of these 210 galaxies,
173 have redshifts in the [0.18,0.24] interval.  The positions of the
210 galaxies with measured redshifts and magnitudes are displayed in
Fig.~\ref{fig:xy_z}. It is interesting to note that, as traced by
these galaxies, both clusters appear to be strongly elongated, with
elongation position angles that differ from one cluster to the
other. This strongly suggests that the clusters are not in dynamical
equilibrium, in agreement with previous studies (see above, and also
Zabludoff et al. 1995, Roettiger et al. 1996, Boschin et al. 2004,
Girardi et al. 2006).

The corresponding redshift histogram, zoomed on the cluster redshift
interval, is shown in Fig.~\ref{fig:histoz}.

\begin{figure} 
\centering \mbox{\psfig{figure=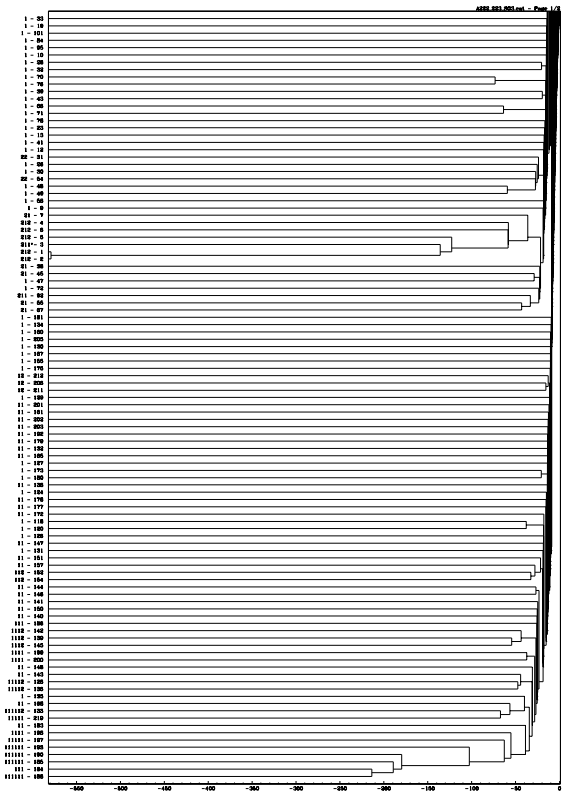,width=8cm}}
\caption[]{Dendogram showing how Abell~222 and Abell~223 can be
  dynamically separated when applying the Serna \& Gerbal
  algorithm. The abscissa is the negative binding energy while the
  catalogue numbers of the various galaxies are shown along the
  ordinate (the top of the dendogram is truncated, but would only show
  a constant gravitational binding energy). }
\label{fig:dendo}
\end{figure}

We applied to the catalogue of 210 galaxies with redshifts and
magnitudes the Serna \& Gerbal (1996) algorithm (hereafter SG), which
allows to search for substructures, separate galaxies forming
gravitationally bound structures, and estimate the masses of these
substructures.  The dendogram thus obtained is shown in
Fig.~\ref{fig:dendo}.  It clearly shows two dynamically distinct
substructures, corresponding to the two clusters: Abell~222 in the top
half and Abell~223 in the bottom half.  The two main gravitationally
bound structures corresponding to these two clusters respectively
include 55 and 64 group members.  Their corresponding velocity
dispersions are 1014 and 1170~km~s$^{-1}$ respectively.  The virial
masses derived for these two substructures (assuming M/L=400 solar
units) are $1.9\ 10^{11}$ and $4.2\ 10^{13}$~M$_\odot$. 

For Abell~222, and probably also for Abell~223, these masses are
obviously underestimated, particularly for Abell~222, since such a
value does not even reach that of a cD galaxy.  We have tried to
analyze the two clusters separately, but this modifies the resulting
masses only slightly. The fact that we find too small cluster masses
can be due to several reasons. First, the redshift sample we are using
is taken from the NED data base, and is obviously not complete, even
at bright magnitudes. We have estimated the completeness of the
redshift catalogue by comparing the numbers of galaxies with redshifts
in the [0.18,0.24] redshift interval to the total number of galaxies,
in magnitude bins of 0.5 in the $r'$ band.  We find that for
$17<r'<19.5$ the completeness is around 50\%, then rapidly decreases
for $r'>19.5$, confirming that the cluster masses based on such a
redshift catalogue are probably unreliable.  Second, as pointed out by
the referee, the SG method relies on the unlikely assumption that
galaxies in clusters are the mass carriers, and additionally requires
an assumption about the M/L ratio, here taken to be M/L=400; this
value may be reasonable for a cluster taken as a whole, but is too
large for individual galaxies (the dark matter halos of individual
galaxies would be overlapping in the dense central regions of
clusters, and strong mass segregation would result by dynamical
friction, which is not observed).

Better estimates of the cluster masses can be obtained by applying
other methods.  For example, the relation between mass and velocity
dispersion computed by Biviano et al. (2006) leads to masses of $1.2\
10^{15}$ and $1.4\ 10^{15}$~M$_\odot$ (with H$_0=70$
km~s$^{-1}$~Mpc$^{-1}$) for Abell~222 and Abell~223 respectively,
based on respective velocity dispersions of 1014 and 1070
km~s$^{-1}$. Note however that since the clusters are not virialized
these values are probably overestimates. For example, Takizawa et
al. (2010) have recently shown from numerical simulations that for
merging systems where the subcluster has a mass larger than one fourth
that of the large cluster, the virial mass can be larger than twice
the real mass. This most probably applies at least to the Abell~223
cluster.

We can also apply the mass-temperature relation estimated from X-ray
data by Arnaud et al. (2005), with the overall temperatures given in
Table 5; this relation gives masses M$_{200}$ of $3.8\ 10^{14}$ and
$4.7\ 10^{14}$~M$_\odot$ for Abell~222 and Abell~223 respectively
(also with H$_0=70$ km~s$^{-1}$~Mpc$^{-1}$). These values may be
closer to the real values than those estimated from the galaxy
velocity dispersions, since e.g. Takizawa et al. (2010) found that for
merging clusters X-ray derived masses were usually more reliable than
virial mass estimations.  However, as explained for example by Nagai
et al. (2007), although the total ICM mass can be measured quite
accurately (to better than $\sim$6\%) in clusters, the hydrostatic
estimate of the gravitationally bound mass is biased low by about
5--20\% throughout the virial region, primarily due to additional
pressure support provided by subsonic bulk motions in the ICM,
ubiquitous in our simulations even in relaxed systems.  Another source
of error is the fact that the cluster temperatures, particularly that
of Abell~223 (see Fig.~\ref{map_kT}) are not homogeneous.  For
example, Rasia et al. (2004) have shown that an incomplete
thermalisation of the gas due to an incomplete virialisation of the
cluster could lead to an underestimate of the mass.  On the other
hand, if the X-ray gas is heated by a collision, the mass could be
overestimated.

We can also note that Mamon (1993) has calculated the bias on the
$M/L$ ratio when the virial theorem is wrongly applied to a group
which is still evolving dynamically (see his figure~11). He found that
in some cases the mass could be underestimated by a factor reaching
100. Therefore the masses estimated for our clusters assuming that
they are virialized are clearly not valid.

Unfortunately, the SG method does not allow to separate the two
substructures in Abell~223 discovered by Dietrich et al. (2002) based
on galaxy density contours (see their figure~6).

Note that Abell~223 is probably close to the intersection of two or
more filaments, since it is massive and substructured, and its GLF is
comparable of that e.g. of Coma (see Section~3.3). On the other hand,
Abell~222 is probably only following its path along a filament.

\subsection{Large-scale structure}

\begin{figure} 
\centering \mbox{\psfig{figure=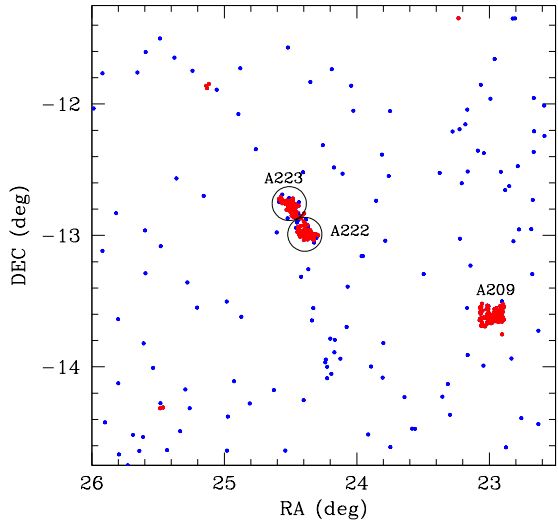,width=8cm}}
\caption[]{Distribution of the galaxies with available redshifts in a
  large region surrounding the clusters. Blue points: all galaxies
  with measured redshifts, red points: galaxies with redshifts in the
  [0.18,0.24] interval. The two clusters under study are circled. A
  third cluster appears to the southwest: Abell~209. The physical size
  of this map is $44.7 \times 44.7$~Mpc$^2$.}
\label{fig:xy_lss}
\end{figure}

Out of the 980 galaxies found in NED in a 3~deg radius region, 310
have redshifts in the [0.18,0.24] interval (i.e. about $\pm
9000$~km~s$^{-1}$ around the mean cluster velocities). Their spatial
distribution is shown in Fig.~\ref{fig:xy_lss}. A cluster is visible
towards the southwest: Abell~209. It has a redshift similar to that of
Abell~222 and Abell~223 and is located about 19.2~Mpc away from
Abell~222 in projection on the sky. No significant filamentary
structure is detected towards the Abell~222/223 clusters.

\section{X-ray analysis}

\subsection{Data reduction}

The pair of clusters Abell~222/223 was observed with XMM-Newton in two
different pointings (revolutions 1378 and 1380) with a total exposure
time of 144~ks.  We used data from all EPIC cameras (MOS1, MOS2 and
pn). The data were reduced with the XMM-Newton Science Analysis
System (SAS) v8.0 and calibration database with all updates available
prior to November 2009. The initial data screening was applied using
recommended sets of event patterns, 0$-$12 and 0$-$4 for the MOS and
PN cameras, respectively.

The light curves are not constant and large variations in intensity
are visible (flares). To improve the signal-to-noise ratio we
discarded periods of flares and the cleaned light curves in the energy
range of [1$-$10] keV exhibited stable mean count rates; exposure
times are given in Table~\ref{obsinfo}. We considered events inside
the field of view (FOV) and excluded all bad pixels.

\begin{table}[ht!]
\centering
\caption{ Information on X-ray data}
\begin{tabular}{ccccccc}
\hline
 Name & Filter & \multicolumn{3}{c}{$t_{\rm exp}$ (ks)} & $n_{H}$ (\rm $10^{20}$ cm$^{-2}$)\\
          & &  MOS1 & MOS2 & pn\\
\hline\noalign{\smallskip}
Abell~222 &    Thin1 & 11.814 & 18.005 & 10.226 & 2.26\\
Abell~223 &    Thin1 & 23.479 & 26.069 & 10.612 & 2.26\\
\hline	  
\end{tabular}
\label{obsinfo}
\end{table}

The background was taken into account by extracting MOS1, MOS2 and pn
spectra from the publicly available EPIC blank sky templates of Andy
Read (Read \& Ponman 2003).  The background was normalized using a spectrum obtained
in an annulus (between 12.5$-$14 arcmin) where the cluster emission is
no longer detected. In one of the pointings, the A223 cluster falls
exactly inside the annulus where we determine the background
contribution.  We thus masked this region, avoiding any cluster
contamination.
In Fig.~\ref{SpecBackNorm} we show a comparison between the observed
and Read backgrounds for the three detectors, as well as the residuals
between both determinations.

\begin{figure}[ht!]
\centering
\includegraphics[width=0.4\textwidth]{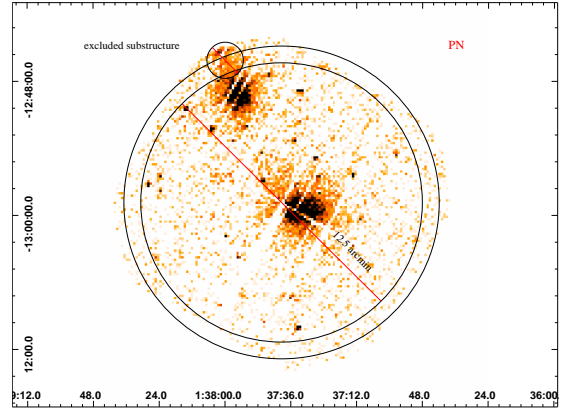}
\caption{Annulus where the background was extracted for
subtraction. }
\label{anel}
\end{figure}

\begin{figure}[h!]
\centering
\includegraphics[width=0.2\textwidth,angle=-90]{Durret_fig16a.ps}
\includegraphics[width=0.2\textwidth,angle=-90]{Durret_fig16b.ps}
\includegraphics[width=0.2\textwidth,angle=-90]{Durret_fig16c.ps}
\caption{Comparison of the observed and Read backgrounds for MOS1
  (\textit{top panel}), MOS2 (\textit{middle panel}) and pn
  (\textit{low panel}). The black points correspond to the observed
  data, the red points to Read's scaled background and the green
  points to the residuals.}
\label{SpecBackNorm}
\end{figure}

\begin{table}[ht!]
\centering
\caption{Overall X-ray temperatures and metallicities}
\begin{tabular}{cccc}
\hline
 Cluster & $r_{200}$ & kT & Z \\
         & (Mpc)      & (keV) & (solar) \\
\hline\noalign{\smallskip}
Abell~222 & $ 1.28 \pm 0.11$ & $3.77 \pm 0.15$ & $0.23 \pm 0.06$\\
Abell~223 & $ 1.55 \pm 0.15$ & $4.38 \pm 0.16$ & $0.23 \pm 0.08 $\\
\hline	  
\end{tabular}
\label{kT_Z}
\end{table}

In order to determine the global properties of these clusters, we
adopted the virial radii $r_{200}$ determined by the weak lensing
analysis (Dietrich et al. 2005).  These values, together with the mean
temperatures and metallicities are given in Table~\ref{kT_Z}.  They
were calculated fixing the hydrogen column density to its galactic
value of $n_{\rm H}=1.56 \times 10^{20} \rm \ cm^{-2}$.

\subsection{Spectrally measured 2D X-ray maps} 
\label{Xrayana}

\begin{figure*}[t!]
\centering
\includegraphics[width=0.8\textwidth]{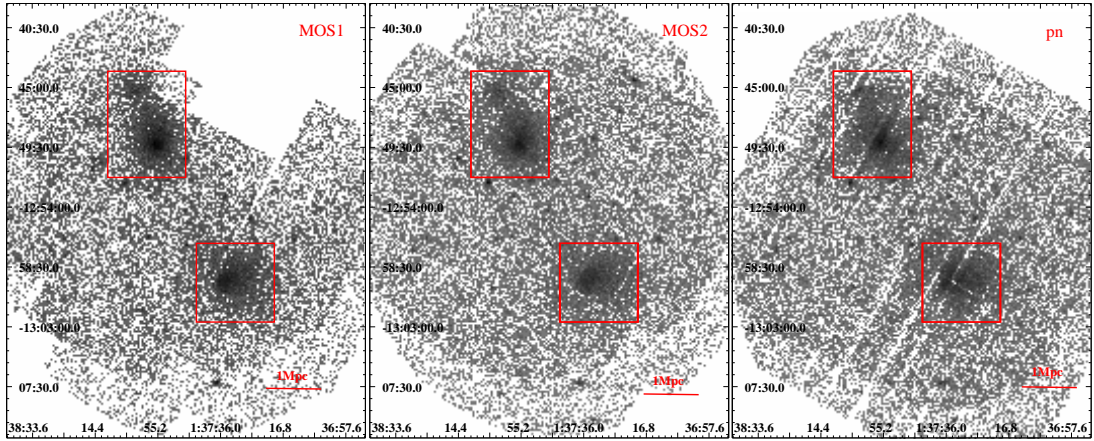}
\caption{Merged event maps from MOS1 (\textit{left panel}), MOS2
  (\textit{central panel}) and pn (\textit{right panel}) showing the
  regions where the 2D maps were obtained.}
\label{RegMapas}
\end{figure*}

We performed quantitative studies using X-ray spectrally measured 2D
maps to derive global properties of these clusters.  These maps were
made in a grid; for each spatial bin we set a minimum count number of
900 (after background subtraction).  For the spectral fits, we used
XSPEC version 11.0.1 (Arnaud 1996) and modeled the obtained spectra
with a MEKAL single temperature plasma emission model (bremsstrahlung
+ line emission, Kaastra \& Mewe 1993, Liedahl et al. 1995).  The free
parameters are the X-ray temperature (kT) and the metal abundance
(metallicity). Spectral fits were made in the energy interval of
[0.7$-$8.0] keV with the hydrogen column density fixed at the
galactic value (see Table~\ref{obsinfo}), estimated with the nH task
of FTOOLS (based on Dickey \& Lockman 1990). 

We compute the effective area files (ARFs) and the response matrices
(RMFs) for each region in the grid.  This procedure (already presented
in Durret et al. 2005 and Lagan\'{a} et al. 2008) allows us to perform
a reliable spectral analysis in each spatial bin, in order to derive
high precision metallicity and temperature maps (see
Fig.~\ref{RegMapas}), since we simultaneously fit all three
instruments for both pointings (that is, we coadd 6 spectra: 2 MOS1, 2
MOS2 and 2 PN).  The best fit value is then attributed to the central
pixel.

\begin{figure}[ht!]
\centering
\includegraphics[width=0.5\textwidth]{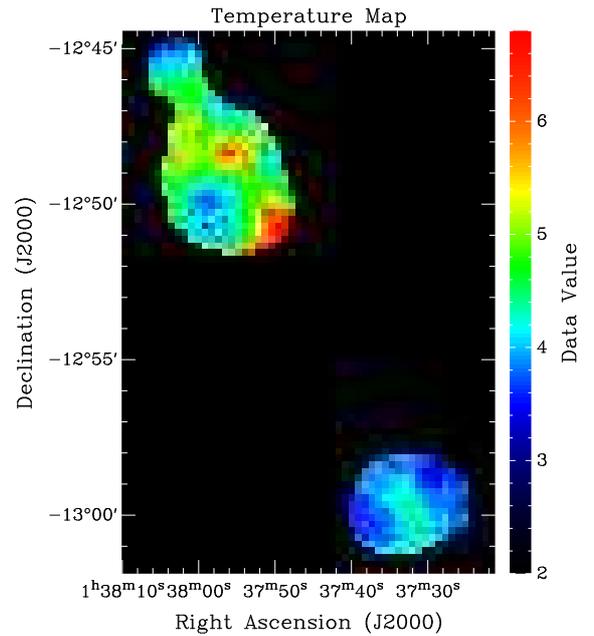}
\caption{X-ray temperature map for Abell~222 (bottom) and Abell~223 (top).
The color-bar indicates the temperature in keV. The corresponding error map 
is shown in Fig.~\ref{err_kT}.}
\label{map_kT}
\end{figure}

\begin{figure}[ht!]
\centering
\includegraphics[width=0.5\textwidth]{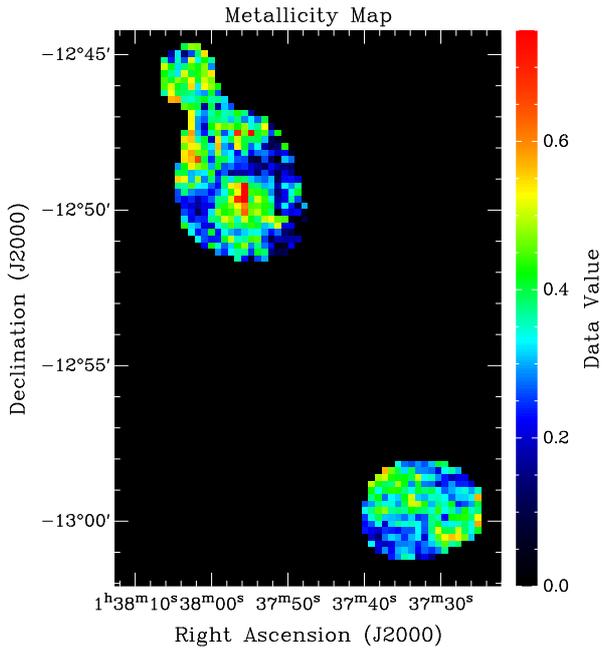}
\caption{X-ray metallicity map for Abell~222 (bottom) and Abell~223
  (top).  The color-bar indicates the metallicity in solar units. The
  corresponding error map is shown in Fig.~\ref{err_Z}.}
\label{map_Z}
\end{figure}

The resulting X-ray temperature and metallicity maps are displayed in
Figs.~\ref{map_kT} and \ref{map_Z}. Maps of the errors on these two
maps are given in the Appendix (Figs.~\ref{err_kT} and
\ref{err_Z}). 

The study of the thermal structure of the intra-cluster medium (ICM) provides a very
interesting record of the dynamical processes that clusters of galaxies
have experienced during their formation and evolution.  The
temperature (and pressure) distribution of the ICM gives us important
insight into the process of galaxy cluster merging and on the
dissipation of the merger energy in form of turbulent motion.
Metallicity maps can be regarded as a record of the integral yield of
all the different stars that have released their metals through
supernova explosions or winds during cluster evolution. 

From the temperature maps we can see that neither cluster presents a 
defined cool-core as a sign of
relaxation. However, Abell~222 is fairly isothermal when compared to
Abell~223.  The overall temperature structure in Abell~223 shows the
clear imprint of the recent interaction(s) this cluster has been
through.  The cluster exhibits a remarkable blob of hotter gas at its
southwest extremity, a colder region to the southeast, and a hotter
zone crossing the cluster along a roughly east-west direction.  The
metallicity map of Abell~222 is quite homogeneous, while that of
Abell~223 is very inhomogeneous, with a metallicity enhancement
in a zone which does not completely coincide with the colder southeast
region and a low metallicity region extending along a zone curving
from the east to the west-northwest.

\subsection{Interpretation} 

As has been shown by Kapferer et al. (2006, see his figure 10), the
enhancement of the metallicity from Z=0.2$Z_{\odot}$ (in the region of
the shock) to Z=0.4$Z_{\odot}$ (in the surrounding region) takes 3~Gyr
(assuming their model B).  This means that the gas has not yet had
time to mix and distribute the metals in the region where the shock is
still propagating. As can be seen in the larger variations in
metallicity that Abell~223 presents when compared to Abell~222, the
metallicity of merging systems is not smoothly distributed. It is
likely that Abell~223 has recently been crossed by a smaller cluster,
which now appears at the north east tip of the maps.

From the metallicity map we also observe that the central region of
Abell~223 presents a metallicity enhancement when compared to the
surrounding medium, though not highly centrally peaked.  At this
point, it is important to highlight that Abell~223 has more very
bright galaxies than Abell~222.  
The results of a former work (Lagan\'a et al. 2009) suggested that
BCGs alone cannot produce the observed metallicity excess in the
central region (De Grandi \& Molendi 2001, De Grandi et al. 2004).
Thus, it is likely that the cool-core has been disrupted by the
sub-cluster collision but the metals injected into the ICM by very
bright galaxies of Abell~223 have not moved that much away from the
central region yet. This can occur if one assumes ram-pressure stripping
with tidal disruption of these very bright galaxies near the center as
a mechanism for metal injection. In this case, we would expect
Abell~223 to have a central metallicity higher than Abell~222, and
this is indeed the case.

\begin{figure}[ht!]
\centering
\includegraphics[width=0.4\textwidth]{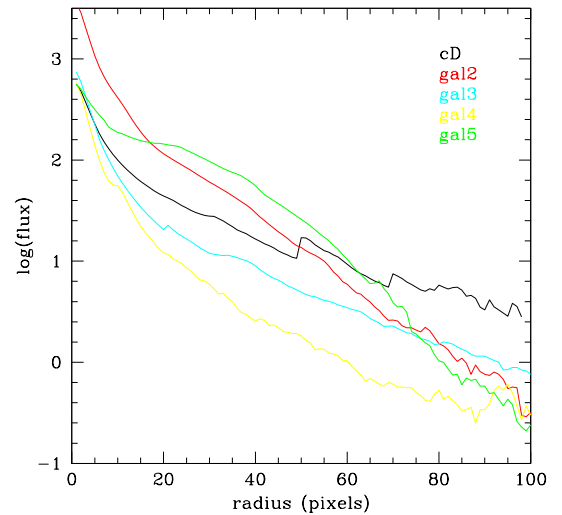}
\caption{Surface brightness profiles of the five brightest galaxies in
  Abell~223. The ``cD'' is in black, the very bright galaxy near the
  cD is in red, and the three other galaxies are somewhat more distant
  from the cluster center. The cD profile decreases notably slower
  with radius that the profiles of the other galaxies.}
\label{fig:profils}
\end{figure}

It is interesting to note that Abell~223 has in fact two BCGs. The
galaxy that we consider as being in the cluster center looks like a
cD, with several smaller satellite galaxies. The surface brightness
profiles of the brightest galaxies of Abell~223 are displayed in
Fig.~\ref{fig:profils}. We can see that the surface brightness profile
of the ``cD'' decreases slowlier than those of the other bright
galaxies.  It can therefore be considered as a dominant galaxy, since
it also shows clear signs of an associated population of fainter
galaxies in the likely process of being accreted. A second nearby
elliptical galaxy which is even brighter could be the central galaxy
of an accreted group. This is reminiscent of the Coma cluster where
the ``cD'' (NGC~4874) is not quite as bright as the brightest galaxy
(NGC~4889), which is believed to have been accreted after the cluster
was formed.

In the case of Abell~222, the situation is even more complicated,
since there are three brightest cluster galaxies, aligned along the
direction seen in Fig.~\ref{fig:xy_z} and having exactly the same
magnitude, the middle one being surrounded by several small satellite
galaxies that give it the appearance of a cD. This is another argument
in favour of the idea that Abell~222 is not fully relaxed either.

\section{Discussion and Conclusions}

Understanding the interplay between galaxies and ICM is fundamental for
further investigations of galaxy clusters. In this way, an important
step is made with multiwavelength studies of these structures.

In order to understand better the connection between ICM and galaxies
in merging systems we combined both X-ray and optical data to examine
in detail the pair of clusters Abell~222/Abell~223. To do so, we
combined spectroscopically computed 2D-maps with optical data to gain
insight on the cluster history.  We verified that besides their
proximity, these clusters show very different evolutionary stages that
can be detected in optical and X-ray analyses. The Abell~223 data
indicate that this cluster is dynamically perturbed when compared to
its companion. Signs of recent merger(s) are detected at both
wavelengths, meaning that the gas and the galaxies are still out of
equilibrium. It is more likely that the signs of merger detected in
Abell~223 are due to the interaction with the subcluster.  On the
other hand, Abell~222 is a smaller cluster closer to a stable
dynamical state than Abell~223.

As mentioned in Sect.~\ref{Xrayana}, Abell~223 exhibits a remarkable
blob of hotter gas at its southwest extremity. The gas of this region
may have been compressed and heated by a recent shock that this
cluster has been through. Another zone of higher temperature is
elongated towards the north east-south western direction.  This is
close to the direction that connects the two clusters and where a
filament of low temperature was already detected between these
clusters (Werner et al. 2008). This region does not appear in our 2D
maps because there are not enough X-ray counts. However, the X-ray
contours overlaid in the 2D maps show that the emission goes beyond
the limits of the maps and towards the region where the filament is
located.

The presence of only a small number of hot blobs and the low X-ray
temperature of the filament tend to indicate that this system is in
the first stages of a merger.  This hints at a scenario in which 
  the two clusters are at the first stages of a collision.  These
clusters are beginning to collide and from the optical counterpart
analysis, the central region of Abell~223 already exhibits signs of
this interaction presenting a higher number of star forming galaxies.
So it seems probable that a small cluster has just crossed Abell~223
and is heading towards the north east.  On the other hand, Abell~222
does not present strong evidence of interaction other than the absence
of a cool-core.

We thus summarize our main results below: 
\begin{itemize}
\item{From the optical analysis we conclude that Abell~222 is smaller
    and less massive than Abell~223. For Abell~222, the GLF is quite
    well fit by a Schechter function in both the $r'$ and $g'$
    bands. On the other hand, since Abell~223 is a double cluster, it
    presents a ``perturbed'' GLF. Abell~223 has more bright galaxies
    than Abell~222 and as a consequence a second component is required
    to fit the GLF at bright magnitudes. The GLFs of both clusters are
    comparable in the $r'$ band but in the $g'$ band the GLF is
    notably steeper in Abell~223 than in Abell~222. This is explained
    by a higher number of faint star forming galaxies present in
    Abell~223. Star formation is triggered by recent merger events.
    Such mergers disturbed the gas and its imprints can be observed in
    2D X-ray maps.}

\item{Both clusters appear to be elongated when traced by their
    cluster galaxy distributions. This is another evidence that these
    clusters are not in equilibrium.  However, 
    the X-ray filament joining the two clusters does not seem to play
    a very important role in the evolution of cluster galaxies. This
    is probably explained by the low temperature of the gas, since
    ram-pressure is not very efficient below 2~keV.}

\item{From the spectrally measured 2D X-ray maps we conclude that
    Abell~222 is fairly isothermal when compared to
    Abell~223. However, neither of these clusters presents a cool-core
    as a sign of relaxation. The overall temperature map of Abell~223
    is an imprint of the recent dynamical interactions this cluster
    has experienced.  The broad variation in temperature puts in
    evidence a large number of sub-structures.  From the metallicity
    map, we also observe an enhancement in the central region of
    Abell~223. This can be explained by the fact that this cluster has
    more very bright galaxies than Abell~222. Assuming ram-pressure
    stripping with tidal disruption of the bright galaxies near the
    cluster center as a mechanism for metal injection, one would
    expect this cluster to have a higher central metallicity than
    Abell~222. The combined analysis of X-ray and optical data argues
    for a scenario in which the two clusters are beginning to collide.
    The former cluster presents signs of interaction in both
    wavelength ranges, and has probably been crossed by a small
    cluster heading north east, while its companion does not show
    strong evidence for interaction, except for the absence of a
    cool-core.}

\end{itemize}

\begin{acknowledgements}

  We are grateful to G.~B. Lima Neto for his help in obtaining the
  X-ray maps.  We warmly thank Andrea Biviano for giving us his
  Schechter function fitting programme free of charge and helping us
  with the corresponding plots. We also wish to thank Delphine Russeil
  for pointing the fact that the clusters under study are in the
  direction of the Sagittarius stream, and Stephen Gwyn for email
  exchanges concerning the Megapipe archive. Last but not least, we
  warmly thank the referee, Andrea Biviano, for his relevant questions
  and constructive comments. This work was supported by FAPESP
  (grants: 2006/56213-9, 2008/04318-7).

\end{acknowledgements}

\appendix

\section{Sectors for Fig.~\ref{fig:ntypesec} }

\begin{figure}[h!]
\centering
\includegraphics[width=0.4\textwidth]{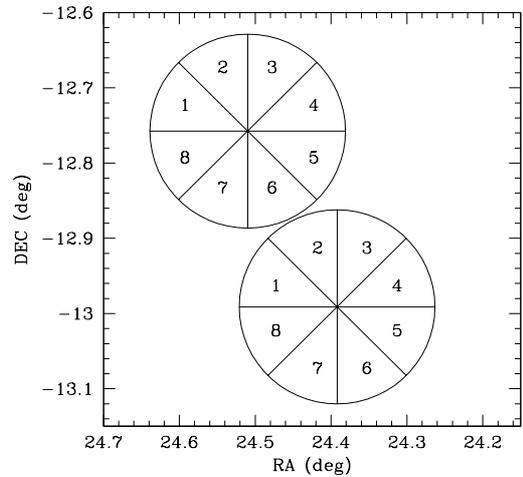}
\caption{Positions of the sectors used to draw Fig.~\ref{fig:ntypesec}.}
\label{secteurs}
\end{figure}

The positions of the sectors used to draw Fig.~\ref{fig:ntypesec}
are shown in Fig.~\ref{secteurs}.

\section{Error maps}

\begin{figure}[t!]
\centering
\includegraphics[width=0.55\textwidth]{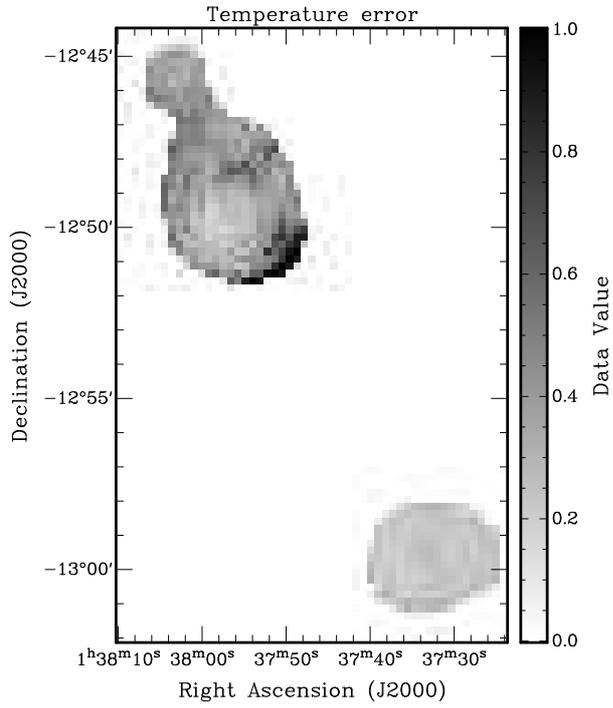}
\caption{Error map on the X-ray temperature.}
\label{err_kT}
\end{figure}

\begin{figure}[t!]
\centering
\includegraphics[width=0.55\textwidth]{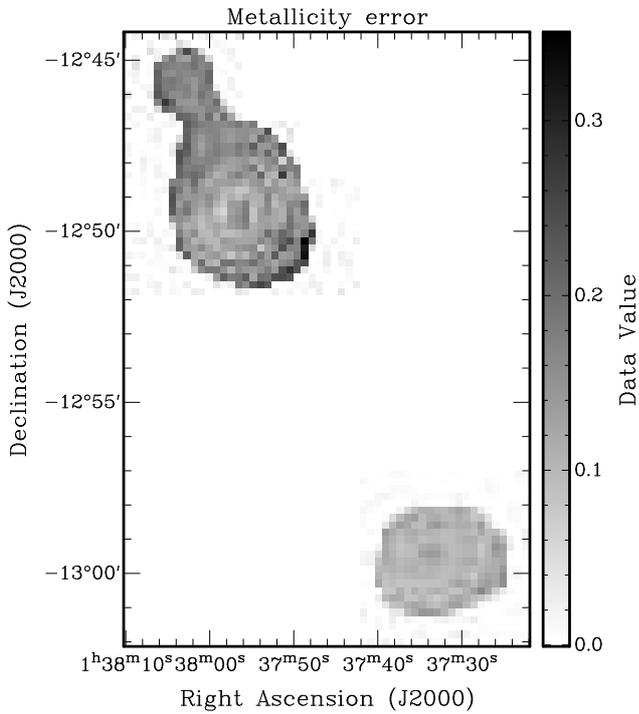}
\caption{Error map on the X-ray metallicity.}
\label{err_Z}
\end{figure}

The maps of the errors on the X-ray temperature and metallicity maps
are displayed in Figs.~\ref{err_kT} and \ref{err_Z}. The errors on these
parameters were directly obtained from the spectral fits and they give the 
deviations in the maps presented in Figs.~\ref{map_kT} and \ref{map_Z}.

\end{document}